\documentclass[letterpaper,11pt,review]{article}
\usepackage{aop}
\usepackage[pdftex,colorlinks=true,bookmarks=false,citecolor=blue,filecolor=blue,urlcolor=blue,pdfstartview=FitH]{hyperref}

\usepackage{graphicx}

\newcommand{\tkmu}{{\tilde k}_\mu}
\newcommand{\eps}{\epsilon}
\newcommand{\bx}{{\bf x}}
\newcommand{\be}{\begin{equation}}
\newcommand{\ee}{\end{equation}}
\newcommand{\beq}{\begin{eqnarray}}
\newcommand{\eeq}{\end{eqnarray}}
\newcommand{\ba}{\begin{array}}
\newcommand{\ea}{\end{array}}
\newcommand{\bea}{\begin{eqnarray}}
\newcommand{\eea}{\end{eqnarray}}

\newcommand{\g}{\gamma_{\perp}}
\newcommand{\gp}{\gamma_{\|}}
\newcommand{\om}{\omega}

\newcommand{\Ep}{E^+}
\newcommand{\Pp}{P^+}

\newcommand{\psimu}{\Psi_\mu(\bx)}

\newcommand{\bxp}{{\bf x^\prime}}

\begin{document}

\title{Modes of Random Lasers}

\author{J. Andreasen$^1$, A. A. Asatryan$^2$, L. C.  Botten$^2$, M.
A. Byrne$^2$, H. Cao$^1$, L. Ge$^1$, L. Labont\'{e}$^3$,  P. Sebbah$^{3,*}$, A. D.
Stone$^1$, H. E. T\"ureci$^4$, C. Vanneste$^3$}

\address{$^1$ Department of Applied Physics, P.O. Box 208284, Yale University, New Haven, CT 06520-8284, USA.}
\address{$^2$ Department of Mathematical Sciences and Center for Ultrahigh-bandwidth Devices for Optical Systems (CUDOS), University of Technology, Sydney, NSW 2007, Australia}
\address{$^3$ Laboratoire de Physique de la Mati\`{e}re Condens\'{e}e, CNRS UMR 6622, Universit\'{e} de Nice-Sophia Antipolis, Parc Valrose, 06108 Nice Cedex 02, France }
\address{$^4$ Institute for Quantum Electronics, ETH Zurich, 8093 Zurich, Switzerland.}

\address{$^*$Corresponding author: sebbah@unice.fr}

\begin{abstract}
In conventional lasers, the optical cavity that confines the photons also determines essential characteristics of the lasing modes such as wavelength, emission pattern, directivity, and polarization. In random lasers, which do not have mirrors or a well-defined cavity, light is confined within the gain medium by means of multiple scattering. The sharp peaks in the emission spectra of semiconductor powders, first observed in 1999, has therefore lead to an intense debate about the nature of the lasing modes in these so-called lasers with resonant feedback. In this paper, we review numerical and theoretical studies aimed at clarifying the nature of the lasing modes in disordered scattering systems with gain. The last decade has witnessed the emergence of the idea that even the low-Q resonances of such open systems could play a role similar to the cavity modes of a conventional laser and produce sharp lasing peaks. We will focus here on the near-threshold single-mode lasing regime where nonlinear effects associated with gain saturation and mode competition can be neglected.  We will discuss in particular the link between random laser modes near threshold and the resonances or quasi-bound (QB) states of the passive system without gain. For random lasers in the localized (strong scattering) regime, QB states and threshold lasing modes (TLM) were found to be nearly identical {\it within} the scattering medium. These studies were later extended to the case of more lossy systems such as random systems in the diffusive regime where it was observed that increasing the openness of such systems eventually resulted in measurable and increasing differences between quasi-bound states and lasing modes. Very recently, a theory able to treat lasers with arbitrarily complex and open cavities such as random lasers established that the threshold lasing modes are in fact distinct from QB states of the passive system and are better described in terms of a new class of states, the so-called constant-flux states. The correspondence between QB states and lasing modes is found to improve in the strong scattering limit, confirming the validity of initial work in the strong scattering limit.
\end{abstract}


\tableofcontents
\setcounter{tocdepth}{2}

\section{Introduction}

The investigation of laser action in complex media with disorder has a long history going back to the early days of laser physics (for a review, see \cite{Wiersma00,CaoReview02,CaoReviewWRM03,CaoReview03,WiersmaNaturePhysics08}). Beginning in the mid 1990's there was a resurgence of interest in this topic both for its intrinsic interest and because of a possible relation to the phenomenon of Anderson localization \cite{Anderson58} previously studied mainly in the context of electronic systems. Random lasers are disordered media with gain that do not possess a light-trapping cavity beyond the confinement provided by multiple-scattering from the disorder itself. Hence they are usually extremely open, low finesse lasers. Initially it was unclear whether such systems could produce narrow lasing lines without any well-confined electromagnetic modes, and while initial experimental studies did find strong amplification near the transition frequency determined by the gain medium, discrete lines were not observed \cite{Migus,Lawandy,Alfano}. Subsequent studies in smaller systems with focused pumping did find discrete lasing lines not necessarily located at the center of the gain curve and photon statistics characteristic of gain saturation \cite{Cao98,Cao99,Frolov99,Cao00,Cao01}, demonstrating that in some cases random lasers (RLs) behave very much like conventional multimode lasers except for their relatively high thresholds and their pseudo-random emission patterns in space. The experimental observations of laser peaks have naturally called for the search for a feedback mechanism leading to light trapping within the scattering medium. There is in fact a case where light can be well-confined inside an open disordered medium. Such confinement occurs when the scattering is extremely strong and the system is in the regime of Anderson localization \cite{Lagendijk09}. However, except in quasi-1D geometries \cite{GenackMilnerPRL05}, the vast majority of experiments on RLs do not appear to be in the localized regime, so the question of whether laser action in a diffusive $(L \gg \ell)$ or quasi-ballistic $(L \sim \ell)$ medium has a qualitatively different nature and origin with respect to conventional lasers remained open for some time (here $L$ is the system size and $\ell$ is the optical elastic mean free path).

With the renewed experimental interest in RLs came also a number of attempts to generalize laser theory to describe such a system.  Early on a major distinction was made between conventional lasers which operate on resonant feedback and RLs which at least in some cases were supposed to operate only on non-resonant feedback (NRF) \cite{CaoReview03}. In the case of NRF the light intensity in the laser was described by a diffusion equation with gain but the phase of the light field and hence interference did not play a role.  A key finding is that there is a threshold for amplification when the diffusion length for escape $L_D \sim dL^2/\ell$ became longer than the gain length (here $d=2,3$ is the dimensionality).  The spatial distribution of intensity above threshold would be given by the solution of a diffusion equation.  In this approach there would be no frequency selectivity and the amplified light would be peaked at the gain center. Clearly such a description would be inadequate to describe RLs based on Anderson localized modes, as such modes are localized in space precisely due to destructive interference of diffusing waves arising from multiple scattering.

This question itself is related to a basic question in non-linear optics: can a system, disordered or not, which is so leaky that it has no isolated linear scattering resonances nonetheless have sharp laser lines due to the addition of gain? And if so, how are the modes associated with these lines related to the broad and difficult to observe resonances of the passive cavity? For an open diffusive or quasi-ballistic medium in two or three dimensions the resonance spacing in wavevector will decrease as $\lambda^{d-1}/L^d$, whereas the linewidth will scale as $\ell/L^2$ (diffusive) or as $1/L$ (ballistic).  Therefore (unless $\ell \approx \lambda$ in $d=2$) the disordered passive cavity resonances strongly overlap and cannot be directly observed in linear scattering.

In the search for a feedback mechanism responsible for the sharp laser peaks observed experimentally \cite{Lagendijk07}, different scenarios have been proposed.  As an alternative to the early picture of closed scattering loops, the probability of having ring-shaped resonators with index of refraction larger than average in the diffusive regime was calculated and shown to be substantially increased by disorder correlation due to finite-size scatterers \cite{Shapiro02,ShapiroJOSAB04}.
Another scenario was put forward where spontaneously emitted photons accumulate gain along very long trajectories. This follows the observation of random spikes in the emission spectrum of weakly active scattering systems in single shot experiments \cite{Mujumdar04,Mujumdar07}. These ``lucky photons" accumulate enough gain to activate a new lasing mode with a different wavelength after each excitation shot.
The experimental study of the modal decay rates in microwave experiments leading to the observation of anomalous diffusion has brought forward the existence of longer-lived prelocalized modes in an otherwise diffusive system \cite{ChabanovPRL03}. An experimental indication of the coexistence of extended and localized lasing modes has been presented recently \cite{Fallert09}. It was suggested that these longer lived-modes could be responsible for lasing. However, although they are possibly achieved in some specific situations, those different scenarios cannot explain the whole set of experimental observations

In this paper, we present recent work, both numerical and analytical, which has shown that within semiclassical laser theory, in which the effects of quantum noise are neglected, definite answers to these questions can be given, without resorting to exotic scenarios. Sharp laser lines based on interference (coherent feedback) do exist not only in strongly scattering random lasers where the localized regime is reached \cite{Soukoulis00,Vanneste01,Vanneste02} but also in diffusive random lasers (DRLs) \cite{Vanneste07,Vanneste09}, and even for weak scattering \cite{Cao06}. Numerical studies have shown that they are associated with threshold lasing modes (TLMs), which, inside the cavity, are similar to the resonances or quasi-bound (QB) states of the passive system  (also called quasi-normal modes). The resemblance is excellent in the localized case \cite{Vanneste01,Vanneste02} and deteriorates as scattering is reduced. A new theoretical approach based on a reformulation of the Maxwell-Bloch equations to access the steady state properties of arbitrarily complex and open cavities allows one to calculate the lasing modes in diffusive and even in weakly scattering random lasers ($\ell \sim L$) \cite{Tureci06,Tureci07,Tureci08,Li08,Tureci09}. A major outcome of this approach is the demonstration that although lasing modes and passive modes can be very alike in random systems with moderate openness in agreement with the above numerical results, they feature fundamental differences. Their distinctness increases with the openness of the random system and becomes substantial for weakly scattering systems. Constant-flux (CF) states are introduced which better describe TLMs both inside and outside the scattering medium for any scattering strength. In addition this theoretical approach allows one to study the multimode regime in DRLs and get detailed information about the effects of mode competition through spatial hole-burning, which appear to differ from conventional lasers.

In this last decade, different types of random lasers (semiconductor powders, pi-conjugated polymers, scattering suspension in dyes, random microcavities, dye-doped nematic liquid crystals, random fiber lasers, ...) have been considered in the literature. We will focus throughout this review mostly on two dimensional (2D) random lasers which consist of randomly distributed dielectric nanoparticles as scatterers. This choice makes possible the numerical and theoretical exploration of 2D finite-sized opened samples where transport can be ballistic, diffusive (in contrast to 1D) or localized \cite{PRB}, by adjusting the index contrast between the scatterers and the background medium.

The outline of this review is as follows: In section 2 we review early numerical explorations of localized and diffusive random lasing demonstrating the existence of threshold lasing modes (TLMs) in all regimes. In section 3 we present recent numerical work based on a time-independent model, which indicates the difference between passive cavity resonances and threshold lasing modes, discussing only single-mode random lasing. The following section will explain why in principle, quasi-bound (QB) states cannot describe TLMs. The last section will introduce the concept of constant-flux (CF) states and describe the self-consistent time-independent approach to describe random lasing modes at threshold as well as in the multimode regime.

\section{Early numerical explorations: Time-dependent model}

\subsection{Localized case}
\label{sec:loc}

From a modal point of view, Anderson localization means that for strong disorder, the eigenmodes of the wave equation are spatially localized in a volume of finite size  $2\xi$, where $\xi$ is the localization length. More precisely, they are spatially localized solutions of the Maxwell equations with tails, which decay exponentially from their center, $\xi$ being the decay length. In the case of scattering particles, the value of the localization length is controlled by the index contrast between the particles and the background medium, the size of the particles, the optical wavelength and the amount of disorder. In practice, when finite-size systems in the localization regime are considered, two opposite cases may occur : (1) $\xi>L$ and (2) $\xi<L$ where L is the system size. In the first case, the system is not large enough for the light to be confined by disorder within the volume of the system. In case (2), light is localized since it cannot escape domains larger than $\xi$. More precisely, localized modes are coupled to the boundaries via their exponential tails. The leakage rate of an exponentially localized quasi-bound (QB) state varies as $\exp(-2r/\xi)$ with $r$ the distance to the boundaries \cite{Pinheiro04} . Hence, in sufficiently large systems QB states located far from the boundaries (which constitutes majority of the QB states except a fraction proportional to $\xi/L$) feature a very small leakage i.e. a good quality factor.

In this subsection, we will consider case (2). Localized modes in a disordered scattering system are quite alike the modes of standard optical cavities, such as the Fabry-Perot. \cite{SebbahPhD}. Hence, one can expect that in the presence of gain, the lasing modes in this regime of strong disorder will be close to the localized QB states of the passive system without gain, in the same way as the lasing modes of a conventional cavity are built with the QB states of the passive cavity. In order to verify that this is really the case one must have access to the individual modes of both the passive system and the active system. Experimentally, such a demonstration has not been achieved yet, essentially because the regime of Anderson localization is difficult to reach and to observe in optics. Besides, until recently there was no fully developed theory describing random lasing modes and their relationship with the eigenstates of the passive system. The easiest way to check this conjecture has been to resort to numerical simulations.

Historically, most of the early numerical studies of random lasers were based on the diffusion equation (see references in \cite{CaoReview03}). However, it is not possible to take into account under the diffusion approximation the interference phenomena which are at the heart of Anderson localization. This is why Jiang and Soukoulis \cite{Soukoulis00} proposed to solve the time-dependent Maxwell equations coupled with the population equations of a four-level system \cite{Siegman86}. The populations $N_i, i=1$ to $4$ satisfy the following equations
\begin{eqnarray}
dN_1/dt=N_2/\tau_{21}-W_pN_1\\
dN_2/dt=N_3/\tau_{32}-N_2/\tau_{21}-(\textbf{E}/\hbar\omega_a)d\textbf{P}/dt\\
dN_3/dt=N_4/\tau_{43}-N_3/\tau_{32}+(\textbf{E}/\hbar\omega_a)d\textbf{P}/dt\\
dN_4/dt=-N_3/\tau_{43}+W_pN_1
\end{eqnarray}
where $W_p$ is the rate of an external mechanism which pumps electrons from the fundamental level (1) to the upper level (4). The electrons in level 4 relax quickly with time constant $\tau_{43}$ to level 3. The laser transition occurs from level 3 to level 2 at frequency $\omega_a$. Hence, electrons in level 3 can jump to level 2 either spontaneously with time constant $\tau_{32}$ or through stimulated emission with the rate
$(\textbf{E}/\hbar\omega_a)d\textbf{P}/dt$. $\textbf{E}$ and $\textbf{P}$ are the electric field and the polarization density respectively. Eventually, electrons in level 2 relax quickly with time constant $\tau_{21}$ from level 2 to level 1. In these equations, the populations $N_i$, the electric field \textbf{E} and the polarization density \textbf{P} are functions of the position \textbf{r} and the time $t$.

The polarization obeys the equation
\begin{equation}
d^2 \textbf{P}/dt^2 + \Delta \omega _a d\textbf{P}/dt + \omega _a^2 \textbf{P}=\kappa.\Delta N.\textbf{E}
\end{equation}\label{Eq2}
where $\Delta N=N_2-N_3$ is the population density difference. Amplification takes place when the rate $W_p$ of the external pumping mechanism produces inverted population difference $\Delta N<0$. The line width of the atomic transition is $\Delta\omega_a=1/\tau_{32}+2/T_2$ where the collision time $T_2$ is usually much smaller than the lifetime $\tau_{32}$. The constant $\kappa$ is given by $\kappa=3c^3/2\omega^2_a\tau_{32}$ \cite{Siegman86}.

\begin{figure}[h]
\centering\includegraphics[clip,width=.6\linewidth]{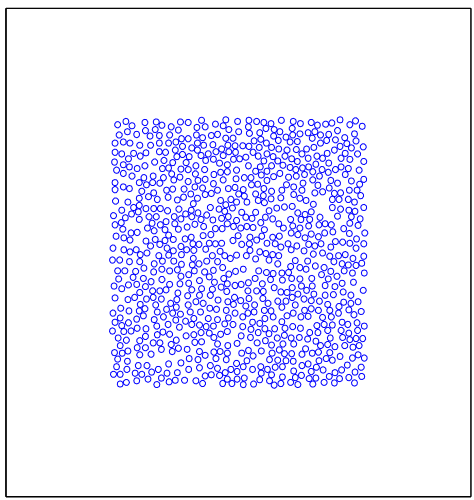}
\caption{(Color online). Example of a random realization of 896 circular scatterers contained in a square box of size L=5 microns and optical index n=1. The radius and the optical index of the scatterers are respectively r=60 nm and n=2 . The total system of size 9 microns is bounded by perfectly matched layers (not shown) in order to simulate an open system}\label{Fig1_0}
\end{figure}

Finally, the polarization is a source term in the Maxwell equations,
\begin{eqnarray}
\partial\textbf{H}/\partial t=-c\nabla\times\textbf{E}\\
\varepsilon(\textbf{r})\partial\textbf{E}\partial
t=c\nabla\times\textbf{H}-4\pi\partial\textbf{P}/\partial t.\label{Eq3}
\end{eqnarray}
The randomness of the system arises from the dielectric constant $\varepsilon(\textbf{r})$, which depends on the position $\textbf{r}$. This time-dependent model has been used in random 1D systems consisting of a random stack of dielectric layers separated by gain media \cite{Soukoulis00} and in random 2D systems consisting of a random collection of circular particles embedded in a gain medium (Fig.~\ref{Fig1_0}) \cite{Vanneste01}. In both cases, a large optical index contrast has been assigned between the scatterers and the background medium to make sure that the regime of Anderson localization was reached. The Maxwell equations are solved using the finite-difference time-domain method (FDTD) \cite{Taflove}. To simulate an open system, perfectly matched layers are introduced at the boundaries of the system \cite{Berenger}. The pumping rate $W_p$ is adjusted just above lasing threshold in order to remain in the single-mode regime.

In 1D, the QB states of the passive system were obtained independently using a time-independent transfer matrix method \cite{Soukoulis02}. In 2D, the Maxwell equations were solved without the polarization term in (\ref{Eq3}) using again the FDTD method. First, the spectrum of eigenfrequencies was obtained by Fourier transform of the impulse response of the system. Next, QB states were excited individually by a monochromatic source at each of the eigenfrequencies.

\begin{figure}[h]
\centering\includegraphics[clip,width=1\linewidth]{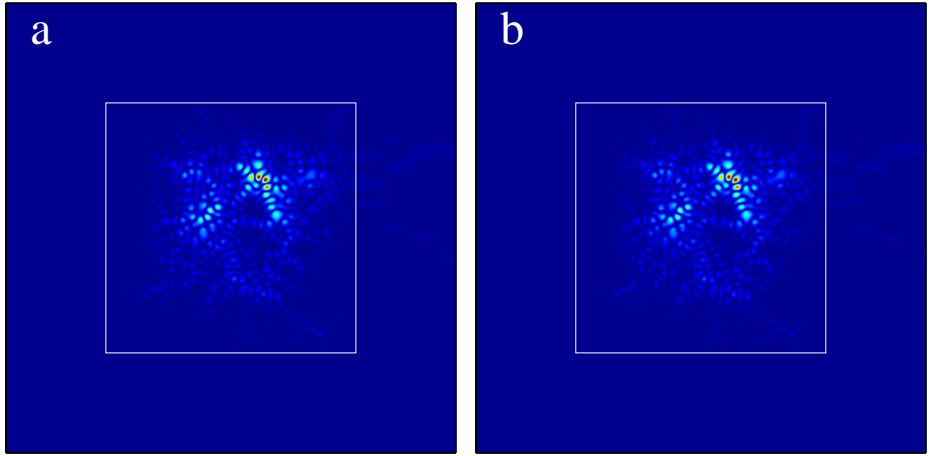}
\caption{(Color online). (a) Spatial distribution of the amplitude of a lasing mode in the localized regime ($n=2$) and (b) that of the corresponding QB states of the same random system without gain. The square delimits the scattering medium. The amplitude rather than the intensity is represented for a better display of the small values of the field.} \label{Fig1_1}
\end{figure}

Finally, in 1D systems \cite{Soukoulis02} as well as in 2D systems \cite{Vanneste02}, lasing modes obtained by the full time-dependent model with gain and localized QB states of the corresponding passive system without gain were compared and found to be identical with a good precision. This was verified for all modes obtained by changing the disorder configuration. An example of a 2D lasing mode and the corresponding QB state of the same system (Fig.~\ref{Fig1_0}) without gain are displayed in Fig.~\ref{Fig1_1}. These results confirmed that the QB states of a localized system play a role similar to the eigenmodes of the cavity of a conventional laser. The only difference is the complicated and system-dependent nature of the localized modes as opposed to the well-known modes of a conventional cavity. These results are in good agreement with the theoretical results described in Section \ref{sec:multimode}, which show that inside systems in the localized regime, the single lasing modes just above threshold are close to the high-Q resonances of the passive system.


\subsection{Diffusive case}\label{sec:dif}

We have seen in the previous section that random lasers in the Anderson localization regime should behave like conventional lasers. They should exhibit discrete laser peaks above threshold in agreement with the experimental observations of laser action with resonant feedback. However, subsequent measurements of the mean free path showed that none of the experimental cases that displayed discrete laser peaks were in the localized regime. Instead, they were found to be in the diffusive regime and some even in the quasi-ballistic regime \cite{Cao06}. In such systems, there are no localized modes so that the observation of laser action with resonant feedback has been the subject of much debate.

\begin{figure}[h]
\centering\includegraphics[clip,width=1\linewidth]{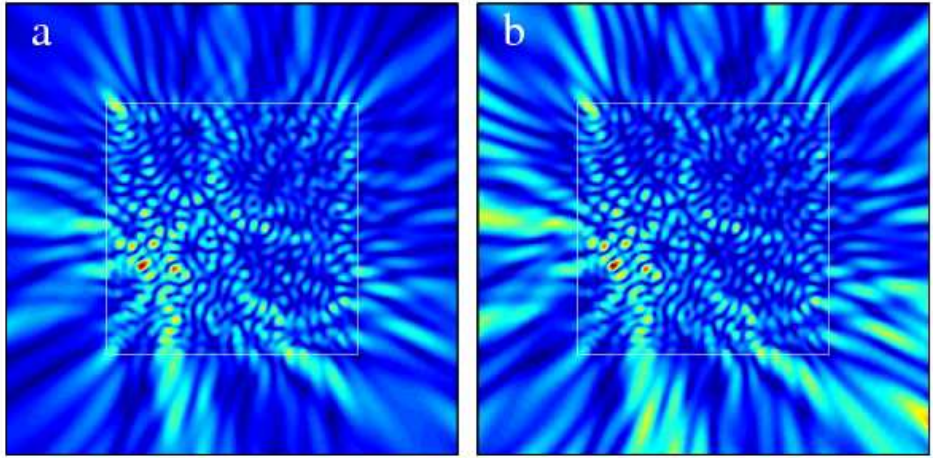}
\caption{(Color online). (a) Spatial distribution of the amplitude of a lasing mode in the diffusive regime. (b) Spatial distribution of the field amplitude after the pump has been stopped and the polarization term has been set to zero. The spatial distribution of scatterers is the collection shown in Fig.~\ref{Fig1_0} but here, the optical index of the scatterers is $n=1.25$ instead of $n=2$ in Fig.~\ref{Fig1_1}} \label{Fig2_2a}
\end{figure}

Only very recently, numerical evidence was given that even diffusive systems with low Q resonances could exhibit lasing with resonant feedback \cite{Vanneste07}. The random 2D systems described in the previous section consisting of random collections of circular particles embedded in a gain medium have been investigated with the same time-dependent model. To be in the diffusive regime instead of the localized regime, a smaller optical index contrast $\delta_n = 0.25$ instead of $\delta_n=1.0$ has been assigned between the scatterers and the background medium. Solving the Maxwell equations coupled to the population equations, laser action characterized by a sharp peak in the emission spectrum was observed just above a threshold, albeit high. An example of the corresponding lasing mode is displayed in Fig.~\ref{Fig2_2a}a. In contrast to the localized case, the lasing mode is now extended over the whole system. Moreover this is a complex mode in the sense that it contains a substantial traveling wave component \cite{Vanneste07}. However, in this work comparison of the lasing modes with the QB states of the passive cavity could not be carried out by using the time domain method as it was done in the localized regime. Due to strong leakage through the boundaries, resonances are strongly overlapping in the frequency domain and one cannot excite them individually by a monochromatic source.

\begin{figure}[h]
\centering\includegraphics[clip,width=.8\linewidth]{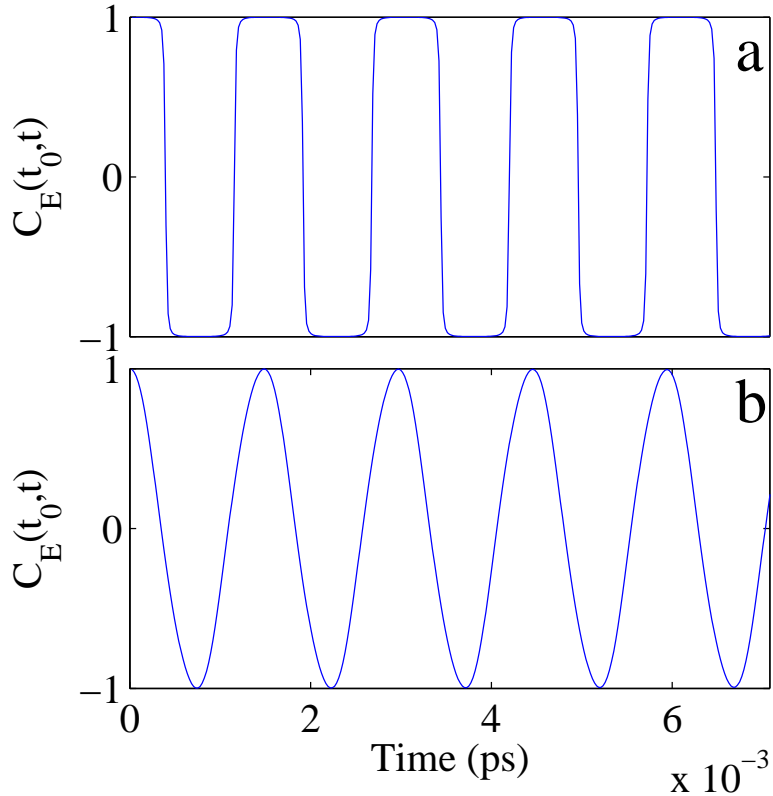}
\caption{(Color online). Short time behavior over a few cycles of the correlation function, $C_{\mathcal{E}}(t_0,t)$, for (a) a localized lasing mode as in Fig.~\ref{Fig1_1} and for (b) a diffusive lasing mode as in Fig.~\ref{Fig2_2a}. The periodic square function in (a) is typical of a standing wave while the sinus-like function in (b) is characteristics of a traveling wave. \cite{Vanneste07}} \label{Fig2_2c}
\end{figure}
\begin{figure}[h]
\centering\includegraphics[clip,width=.8\linewidth]{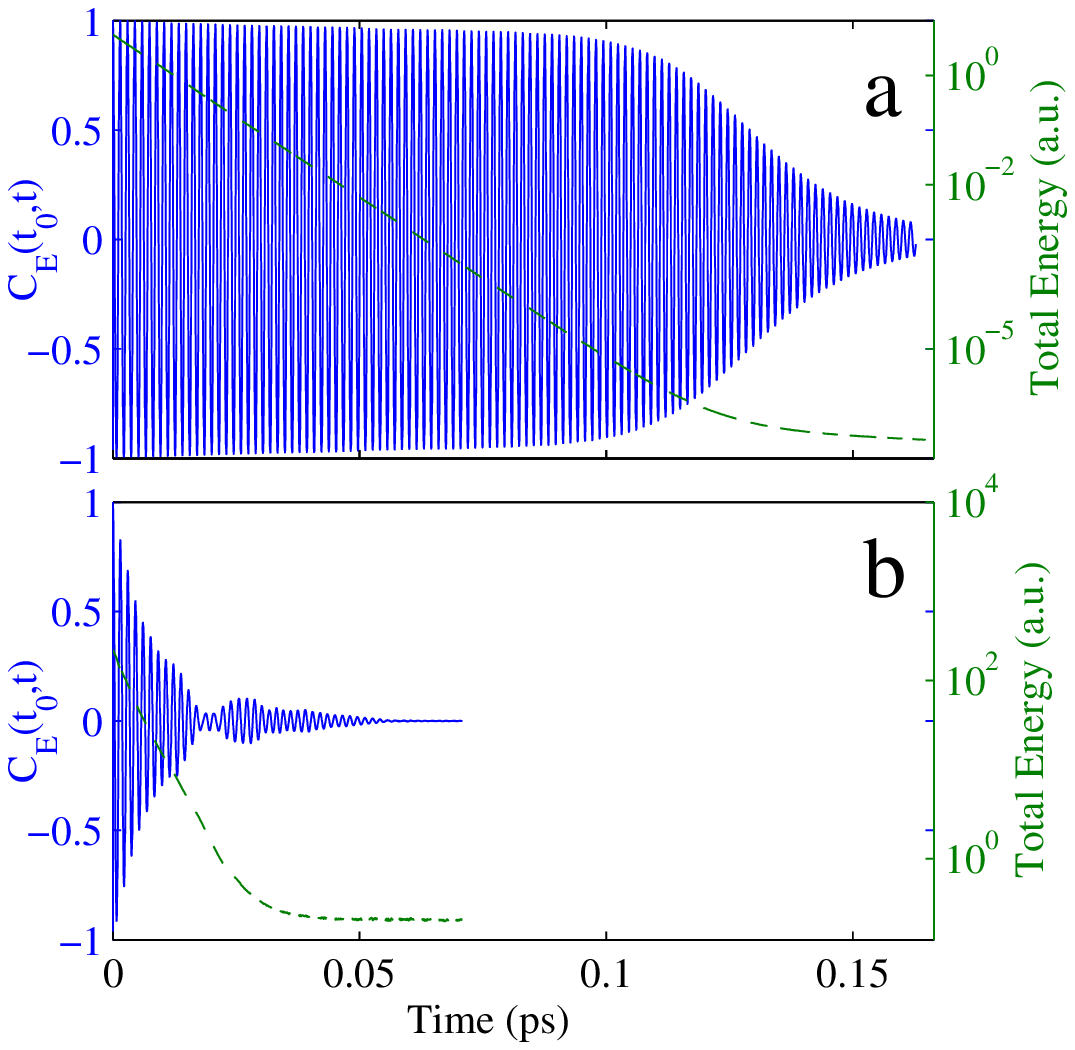}
\caption{(Color online). Correlation function (full line) and energy decay (dashed line) vs time of (a) the lasing mode when the pump is turned off and (b) an arbitrary field distribution at the frequency of the lasing mode} \label{Fig2_2b}
\end{figure}

To circumvent this difficulty, an indirect method has been used to compare the lasing modes with the resonances of the passive system. This method is inspired by the Fox-Li modes, which in conventional laser physics are modes of an open cavity \cite{FoxLi1960,Siegman86,Dutra00}. The Fox-Li modes are field distributions whose profile is self-repeating in a complete round trip within the Fabry-Perot laser cavity while decaying because of the diffraction losses due to finite surface area of the end mirrors. Analogously, if the lasing modes of the diffuse system are related to the resonances of the passive system, they should decay by self-repeating themselves when pumping and population inversion are turned off. To study the evolution of the mode profile with time, the following spatial correlation function was introduced \cite{Vanneste07}
\begin{equation}
C_{\mathcal{E}}(t_0,t)=\int\int_{\cal D} d^2\vec{r}
\mathcal{E}(\vec{r},t_0)\mathcal{E}(\vec{r},t)
\end{equation}\label{Eq2.2.1}
which compares the mode profile $\mathcal{E}(\vec{r},t)$ at time $t$ with the mode profile at the initial time $t_0$. Here, ${\cal D}$ is the scattering medium. The field has been normalized $\mathcal{E}(\vec{r},t)=E(\vec{r},t)/[\int\int_{\cal D} d^2\vec{r}E^2(\vec{r},t)]^{1/2}$ to counterbalance the decay due to the leakage through the boundaries. This correlation  function oscillates at the laser frequency between -1 and +1 if the normalized mode profile is recovered at each period (Fig.~\ref{Fig2_2c}). Otherwise, the amplitude of the oscillations should decay with time. This correlation function was used in \cite{Vanneste07} to check whether the first lasing mode at threshold for diffusive random laser indeed corresponds to a Fox-Li mode of the passive system. The pumping is set to zero  after the lasing mode has been established so that at later times the field can evolve by itself. The long time evolution of the spatial correlation function associated with this free field is displayed in Fig.~\ref{Fig2_2b}a. The decay of the total energy of the system is also shown. While energy decay is observed over 6 orders of magnitude, the spatial correlation function is seen to oscillate between values close to -1 and +1 meaning that the initial lasing mode profile $\mathcal{E}(\vec{r},t_0)$ is reproduced at each period with a good accuracy. The decaying field amplitude has the spatial distribution which is shown in Fig.~\ref{Fig2_2a}b until eventually, the correlation function decays to zero when the decaying field reaches the noise level. This result demonstrates that the TLM is very close to a resonance of the passive system, when measured \emph{inside} the scattering medium. For comparison, the evolution of the spatial correlation function for an initial field created by an arbitrary distribution of monochromatic sources at the laser frequency is displayed in Fig.~\ref{Fig2_2b}b. The fast decay of $C_{\mathcal{E}}(t_0,t)$ after the sources have been turned off indicates that this field distribution is not a QB state of the passive system.

The decay rate observed corresponds to a quality factor of $30$, to be compared with the value $10^4$ found in the localized case. This result shows that a ``bad" resonance in a leaky disordered system can nevertheless turn into a lasing mode in the presence of an active medium. This result is in stark contrast with the common belief that random lasing with resonant feedback involves the presence of resonances with high quality factors. It provides a consistent explanation for the experimental observation of random lasing with resonant feedback even far from the localized regime, without resorting to other scenarios such as those reviewed in the introduction \cite{Lagendijk07,Shapiro02,ShapiroJOSAB04,Mujumdar04,Mujumdar07}.

The comparison of patterns between Fig.~\ref{Fig2_2a}a and Fig.~\ref{Fig2_2a}b shows that the lasing mode and the QB modes are close to each other inside the scattering system as confirmed by the evolution of the correlation function, which has been defined only inside the system. However, one also notices that outside the scattering medium, the field distributions differ substantially. The free propagating field outside the scattering system in Fig.~\ref{Fig2_2a}b reproduces the laser field distribution in Fig.~\ref{Fig2_2a}a with significant distortions due to the enhancement of the amplitude towards the external boundaries of the total system. Hence, the comparison between both figures indicates that if the lasing modes and the QB modes are similar inside the scattering system, they differ noticeably outside. Moreover, a careful examination of the correlation function in Fig.~\ref{Fig2_2b}a shows that it oscillates between two extremal values, which slowly depart from -1 and +1 well prior to the ultimate fast decay. This is in contrast with the long time behavior of the correlation function in the localized regime (not shown), which displays oscillations between -1 and +1 with a very good precision for time scales much longer than the time scale in Fig.~\ref{Fig2_2b}a. This result indicates also that inside the scattering system, the lasing mode is close to but not identical to a QB state.

In conclusion, the time dependent model has provided direct evidence of the closeness of lasing modes and passive cavity resonances at least in the localized case. In the diffusive regime, the lasing modes are also found rather close to the QB modes although small discrepancies manifest themselves. We also found that this holds inside the scattering medium. Outside the scattering system however, differences become more significant. The advantage of the time dependent model is that one has access in principle to the full non-linear dynamics of the laser system. However QB states with low quality factors are not accessible with this approach. Hence, the measure of the difference between TLM and QB states has been indirectly achieved by using the spatial correlation function. Another limitation of this method is related to the various time constants involved in this model which lead to time consuming computations, particularly when one wishes to vary disorder and study an ensemble of disorder configurations. To overcome these limitations, different approaches such as solving the wave equation in the frequency domain have been used. Several approaches of this kind will be described in the next section \cite{Cao00,Beenakker96,Asatrian98,Soukoulis99}. The recent theoretical approach based on a different class of states, the so-called constant flux (CF) states, and taking into account non-linear interactions will be described in section \ref{sec:multimode}.

\section{Numerical simulations: Time-independent models}

Different models have been proposed in the frequency domain to solve the wave equation. In 1D, it is possible to employ the transfer matrix method similar to that used in \cite{Soukoulis02} for studying the lasing modes in an active layered random system. A direct comparison between TLMs and quasi-bound states of the corresponding passive random system is proposed in the first part of this section. In 2D, the multipole method has been used, which also provides a direct comparison of the QB states and the lasing modes of a 2D-disordered open system. The comparison presented in the second part of this section has been carried out for refractive index of the scatterers ranging from $n'_l=2.0$, (localized regime) to $n'_l=1.25$ (diffusive regime). We alternatively used a different approach based on the finite element method to obtain the passive modes, which turned out to be much less computationally demanding in the weakly scattering regime. A brief description of both methods is provided in Appendices \ref{apA} and \ref{femap}.

\subsection{One-dimensional random lasers}

Employing the transfer matrix method, similar to that used in \cite{Soukoulis02}, we study the lasing modes in a one-dimensional (1D) random system and compare them with the QB states of the passive random system. The random system is composed of $161$ layers. Dielectric material with index of refraction $n_1=1.05$ separated by air gaps ($n_2=1$) results in a spatially modulated index of refraction $n(x)$. Outside the random medium $n_0 = 1$. The system is randomized by specifying thicknesses for each layer as $d_{1,2} = \left<d_{1,2}\right>(1+\eta\zeta)$, where $\left<d_1\right>=100$ nm and $\left<d_2\right>=200$ nm are the average thicknesses of the layers, $\eta = 0.9$ represents the degree of randomness, and $\zeta$ is a random number in (-1,1). The length of the random structure $L$ is normalized to $\left<L\right>=24100$ nm. Linear gain is simulated by appending an imaginary part to the dielectric function $\epsilon(x)=\epsilon'(x)+i\epsilon{''}(x)$, where $\epsilon'(x)=n^2(x)$. This approximation is valid at or below threshold \cite{Soukoulis99}. The complex index of refraction is given by $\tilde{n}(x) = \sqrt{\epsilon(x)} = n'(x) + in{''}$, where $n{''} < 0$.
We consider $n{''}$ to be constant everywhere within the random system. This yields a gain length $l_g=|1/k{''}|=1/|n{''}|k$ ($k=2\pi/\lambda$ is the vacuum frequency of a lasing mode) which is the same in the dielectric layers and the air gaps. The real part of the index of refraction is modified by the imaginary part as $n'(x)=\sqrt{n^2(x)+n{''}^2}$.

\begin{figure}[h]
\centering\includegraphics[clip,width=.8\linewidth]{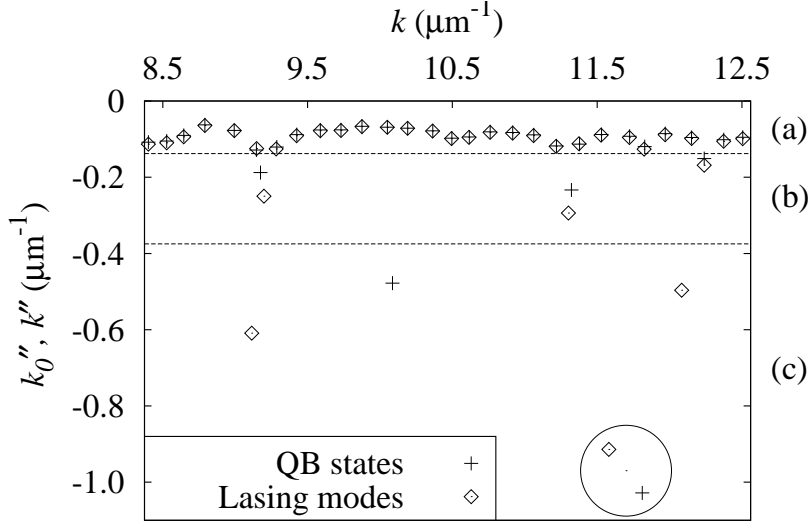}
\caption{The frequencies $k$ of quasi-bound modes (crosses) and lasing modes with linear gain (open diamonds) together with the decay rates $k_0^{''}$ of QB states and the lasing thresholds $k^{''}$ of lasing modes. The horizontal dashed lines separate three different regions of behavior: (a) lasing modes are easily matched to QB states, (b) clear differences appear but matching lasing modes to QB states is still possible, (c) lasing modes have shifted so much it is difficult to match them to QB states. The QB state with the largest decay rate and the lasing mode with the largest threshold are circled, though they may not be a matching pair.}
\label{fig:jahc1}
\end{figure}
We find the frequency $k$ and threshold gain $k^{''}$ of each lasing mode within the wavelength range 500 nm $< \lambda < $ 750 nm. The results are shown in Fig.~\ref{fig:jahc1}. Finding `matching' QB states for lasing modes with large thresholds (large $|k^{''}|$) is challenging due to large shifts of the solution locations
[Fig.~\ref{fig:jahc1}(region c)]. However, there is a clear one-to-one correspondence with QB states for the lasing modes remaining [Figs.~\ref{fig:jahc1}(region a) and (region b)]. It is straightforward to find the matching QB states for these lasing modes and calculate their differences. The average percent difference between QB state frequencies and lasing mode frequencies in Fig.~\ref{fig:jahc1}(region a) is 0.013\% while it is 0.15\% in Fig.~\ref{fig:jahc1}(region b). 
The average percent difference between QB state decay rates $k_0^{''}$ and lasing thresholds $k^{''}$ in Fig.~\ref{fig:jahc1}(region a) is 2.5\% and in Fig.~\ref{fig:jahc1}(region b) is 21\%.

\begin{figure}[h]
\centering\includegraphics[clip,width=.8\linewidth]{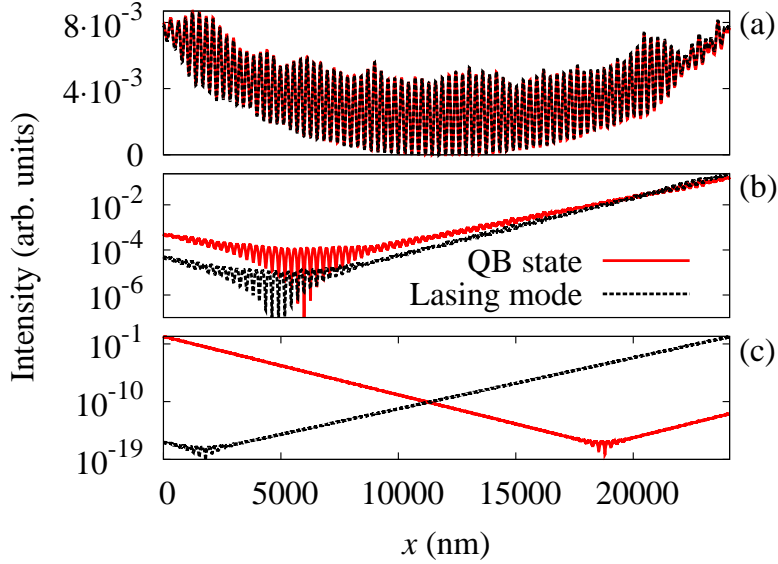}
\caption{(Color online) Spatial intensity distributions of quasi-bound modes $I_{QB}(x)$ (red solid lines) and lasing modes $I_{LG}(x)$ (black dashed lines) from each of the three regions in Fig.~\ref{fig:jahc1}. Representative samples were chosen for each case. (a) The lasing mode intensity is nearly identical to the QB state intensity with $\sigma_d=1.7$\%. (b) A clear difference appears between the lasing mode and the QB state, with $\sigma_d=21.8$\%, but they are still similar. (c) The lasing mode with the largest threshold and QB state with the largest decay rate are compared, with $\sigma_d=198$\%. Though these two modes are fairly close to each other [circled in Fig.~\ref{fig:jahc1}(region c)], their intensity distributions are quite different.}
\label{fig:jahc2}
\end{figure}
The normalized intensities of the QB states $I_{QB}$ and lasing modes with linear gain $I_{LG}$ are also compared. Figure \ref{fig:jahc2} shows representative `pairs' of modes from the three regions shown in Fig.~\ref{fig:jahc1}. The spatially averaged relative difference between each pair of modes is calculated by
\begin{equation}
\langle \sigma_d\rangle=\frac{\int |I_{QB}-I_{LG}|dA}{\int I_{LG}dA}\times 100\%. \label{error}
\end{equation}
For small thresholds [Fig.~\ref{fig:jahc2}(a)], the difference between the lasing modes and the matching QB states is very small. The average percent difference between all pairs of modes in this region is $\left<\sigma_d\right>=4.2$\%. For lasing modes with slightly larger thresholds [Fig.~\ref{fig:jahc2}(b)], there are clear differences. Nevertheless, we may confidently match each lasing mode in this region with its
corresponding QB state. The average percent difference between all pairs of modes in this region is $\left<\sigma_d\right>=24$\%. As mentioned earlier, it is challenging to find matching pairs of lasing modes and QB states for large thresholds. Figure \ref{fig:jahc2}(c) compares the lasing mode with the largest threshold and the QB state with the largest decay rate [circled in Fig.~\ref{fig:jahc1}(region c)]. Though these two modes are fairly close to each other in terms of $k$, $k_0^{''}$, and $k^{''}$, their intensity distributions are quite different. Indeed, there may be no correspondence between the two.

The deviation of the lasing modes from the QB states can be explained by the modification of the transfer matrix. In the passive system, $k_0^{''}$ is constant, but $k^{''}i=k_0^{''}n(x)$ varies spatially. With the introduction of gain, $k^{''}$ becomes constant within the random system, and feedback due to the inhomogeneity of $k^{''}$ is removed. However, introducing gain generates additional feedback inside the random system caused by the modification in the real part of the wave vector $k'=kn'(x)$. Neglecting this effect results in some correspondence between lasing modes and QB states even at large thresholds \cite{Wu07}. Furthermore, since there is no gain outside the random system, $k^{''}$ suddenly drops to zero at the system boundary. This discontinuity of $k^{''}$ generates additional feedback for the lasing modes. In this weakly scattering system, the threshold gain is high. The large drop of $k^{''}$ at the system boundary makes the additional feedback stronger.

\subsection{Two-dimensional random lasers}

We turn now to the 2D case. A different approach based on the multipole method has been used. The multipole method is best suited to characterize multiple scattering problems involving scatterers with circular cross-section.  This method has been used to compute the scattering of a plane wave by a random collection of cylinders \cite{Felbacq,Cao06}, to calculate the defect states in photonic crystals \cite{Centeno99}, to construct the exact Green's function of a finite system \cite{Asatryan03}, or to calculate the local density of states \cite{Asatryan01}. This method has also been used to explain the anomalously large Lamb shift that occurs in photonic crystals by calculating the QB states in such structure \cite{Asatryan06}. Finally, The multipole method can be used to characterize the modes  of three dimensional structures composed of cylinders \cite{McPhedran} and in particular to find the modes of the phothonic crystal fibers \cite{RenversezA,RenversezB,AsatryanPRE03}. It will be used here to calculate the QB states and the lasing modes of the 2D-disordered scattering systems of the kind shown in Fig.~\ref{Fig1_0} and studied in the previous section for different regimes of scattering. Details about this method can be found in Appendix \ref{apA}.

This method is based essentially on a search for the poles of a scattering matrix. Because the system is open, the problem is not Hermitian and hence there are no modes occurring for real wavelengths. The poles of the QB states all occur in the complex plane at wavelengths $\lambda=\lambda'+i\lambda{''}$, with causality requiring that $\lambda{''}>0$. The real part of the wavelength $\lambda'$ determines the resonance wavelength of the QB state, while the imaginary part $\lambda''$ determine the quality factor $Q=\lambda'/(2\lambda'')$ of the mode \cite{Asatryan06}.

The same method is used to find the lasing modes (TLM) at threshold. It is necessary to find this time the poles of the scattering matrix in the two-dimensional space $(\lambda', \varepsilon''_b)$ of real wavelengths ($\lambda''=0$) and imaginary component of the complex dielectric constant outside the scatterers where the gain is distributed. It can also be used to find the lasing modes when gain is localized inside of the scatterers. In this case the poles of the scattering matrix are searched  in the space of real wavelengths ($\lambda''=0$) and the imaginary part of the dielectric constant of cylinders  $\varepsilon''_l)$

The multipole method is both accurate and  efficient: the boundary conditions are analytically satisfied, thus providing enhanced convergence, particularly when the refractive index contrast is high. However, in the case of large systems the method can be slow (given that field expansions are global, rather than local) when it is necessary to locate all poles within a sizable wavelength range. An other extremely efficient time-independent numerical method based on Finite Element Method \cite{Jin93} has been tested. This method is briefly described in the \ref{femap}. We checked that the results obtained by both methods, the (purely numerical) finite element method  and the (semi-analytic) multipole method were identical with a good precision.

\begin{figure}[h]
\centering\includegraphics[clip,width=.8\linewidth]{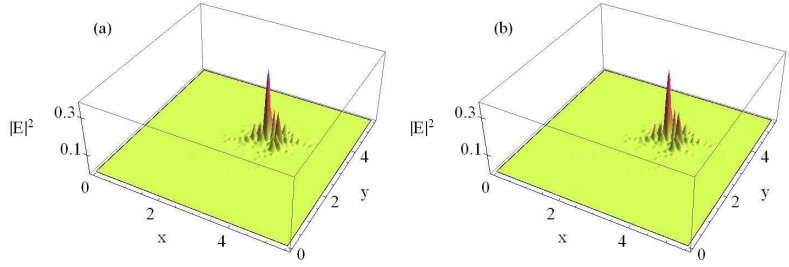}
\caption{(Color online) The intensity $|E|^2$ of the localized QB state (a) (\href{run:Epassive-008-125-topdown.wmv}{Media~1}) and corresponding lasing mode (b) (\href{run:eactive-008-125-topdown.wmv}{Media~2}) calculated using multipole method for a 2D-disordered scattering system of the kind shown in Fig.~\ref{Fig1_0} with the refractive index of the cylinders $n'_l=2.0$.}
\label{Fig2_Ara}
\end{figure}

\subsubsection{Localized case}

We first consider the localized case ($n'_l=2.0$) for which a complete comparison of the QB states and the lasing modes was possible with the time-dependent FDTD-based method (section \ref{sec:loc}), thus providing a reference comparison for the multipole calculations. The lasing mode is found at a wavelength $\lambda'=446.335$~nm for a value of the imaginary part of the refractive index $n''_l=-1.967\times10^{-4}$, representing the pumping threshold for this mode. The spatial distribution of its amplitude is shown in Fig.~\ref{Fig2_Ara}(b). The QB states of the passive system are calculated in the spectral vicinity of the lasing mode. The number of required multipoles was $N_{max}=4$ (see \ref{apA}). Figure \ref{Fig2_Ara}(a) shows the QB state which best resembles the lasing mode. Its wavelength and quality factor are respectively $\lambda'=446.339$~nm and $Q=8047$. The relative difference between the two modes is $\left<\sigma_d\right>=0.05$\%. These calculations provide confirmation that the lasing modes and the QB states are the same inside the scattering region for high $Q$-valued states.
\begin{figure}[h]
\centering\includegraphics[clip,width=.8\linewidth]{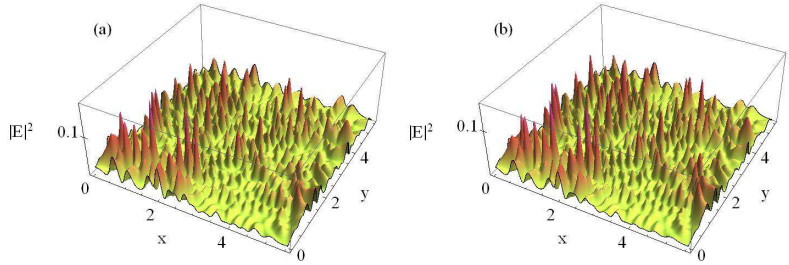}
 \caption{(Color online). The  intensity $|E|^2$ of the diffusive QB state (a) (\href{run:Dpassive-008-125-topdown.wmv}{Media~3}) and the lasing mode (b) (\href{run:Dactive-008-125-topdown.wmv}{Media~4}) calculated using the multipole method for the same random configuration as in Fig.~\ref{Fig2_Ara} but with the refractive index of the cylinders of $n'_l=1.25$.}
  \label{Fig3_Ara}
\end{figure}

\subsubsection{Diffusive case}

We next consider the diffusive case and choose $n'_l=1.25$. This is where the time-independent method becomes interesting since, in contrast to the FDTD approach, it gives a direct access to the QB states. They are accurately calculated in this regime for $N_{max}=2$ multipoles. Figure \ref{Fig3_Ara} shows a lasing mode and its corresponding QB state. The lasing mode is found at $\lambda'=455.827$~nm for an imaginary part of the refractive index $n_l''=-3.778\times 10^{-2}$. The wavelength and the quality factor of the QB state are respectively $\lambda'=456.79$~nm and $Q=29.2$. The lasing mode is therefore red-shifted relative to the QB state's wavelength, as a result of the mode-pulling effect. The QB state and the lasing mode appear similar in Fig.~\ref{Fig3_Ara}. However, the relative difference between the two modes is larger than in the localized case, $\left<\sigma_d\right>=14.5$\%.
\begin{figure}[h]
\centering\includegraphics[clip,width=.7\linewidth]{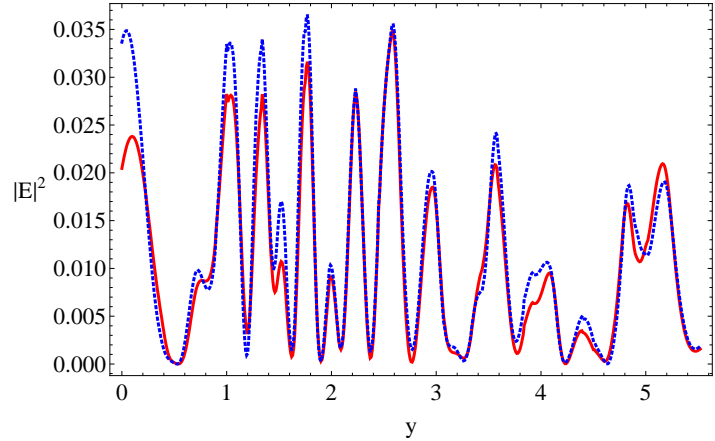}
\caption{ (Color online). The intensity $|E|^2$ of the diffusive QB state (blue dashed line) and lasing mode (red solid line) for $x=2.75$ and $n'_l=1.25$.}
\label{Fig3a_Ara}
\end{figure}
Fig.~\ref{Fig3a_Ara} shows the cross-section of the spatial intensity of both modes along $x=2.75$. In spite of the resemblance, the two profile display visible dissimilarities. This suggests, in the diffusive case, that QB states and lasing modes are not exactly the same though they exhibit quite similar features. These results are consistent with the findings presented in section \ref{sec:dif}.

\subsubsection{Transition case}

\begin{figure}[h]
\centering\includegraphics[clip,width=.8\linewidth]{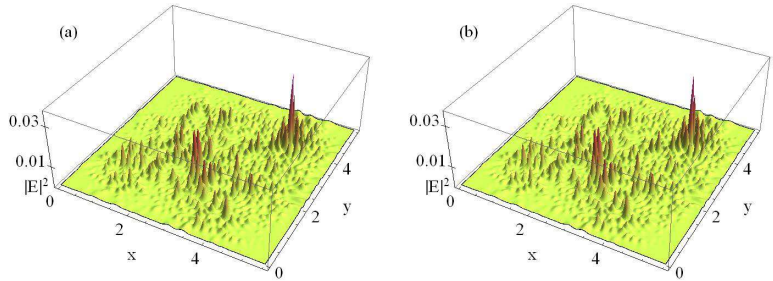}
\caption{(Color online) The intensity $|E|^2$ of a QB state (a) and a lasing mode (b) calculated using multipole method for the same random configuration as above but with the refractive index of the cylinders $n'_l=1.75$.}
\label{Fig4_Ara}
\end{figure}
\begin{figure}[h]
\centering\includegraphics[clip,width=.8\linewidth]{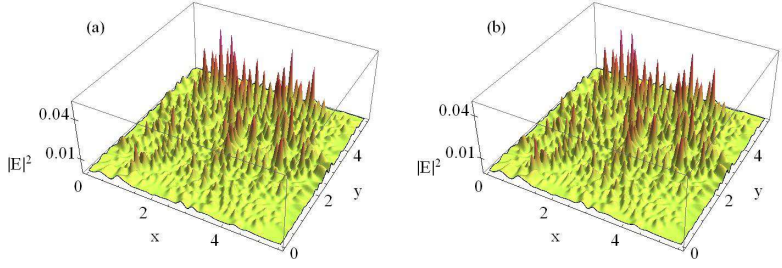}
\caption{(Color online) The same as in Fig.\ref{Fig4_Ara} but for $n'_l=1.5$.}
\label{Fig5_Ara}
\end{figure}
It is both informative and interesting to follow the evolution of the lasing modes and QB states spatial profile when the index of refraction is decreased progressively, allowing to compare systematically the QB states and the random lasing modes (TLM) in regime ranging from localized to diffusive. The QB state and lasing modes calculated for intermediate cylinder refractive indices $n_l'=1.75$ and $n_l'=1.5$ are displayed in Figs. \ref{Fig4_Ara} and \ref{Fig5_Ara}.
We note that the highly spatially localized mode for $n'_l=2$ (Fig.~\ref{Fig2_Ara}) is replaced for $n'_l=1.75$ by a mode formed by two spatially localized peaks and several smaller peaks. For a refractive index of $n'_l=1.5$, the mode is still spatially localized although on a larger area, but is now formed with a large number of overlapping peaks. A more systematic exploration of the nature of the lasing modes at the transition between localized states and extended resonances can be found in \cite{Vanneste09}. There, a scenario for the transition has been proposed based on the existence of necklace states which form chains of localized peaks, resulting from the coupling between localized modes. The modes shown here support this scenario. It is important to note that the decreasing scattering and increasing leakage not only affect the degree of spatial extension of the mode but also the nature of the QB states. Indeed, it was shown in \cite{Vanneste09} that because of leakage, extended QB states have a non-vanishing imaginary part associated with a progressive component, in contrast to the purely stationary localized states. We present in \cite{movie} animations of the time oscillation of the real part of the field $\Re{[\Psi\exp(-i\omega t)]}$ of the QB state and of the corresponding TLM for $n'=2$ and $n'=1.25$. The QB state is exponentially decaying in contrast to the lasing mode. The diffusive lasing mode clearly exhibit a progressive component, which does not exist in the localized lasing mode.

\begin{table}[h]
\centering\begin{tabular}{|c|c|c|c|c|}
   \hline
   $n_l'$ & 2.0 & 1.75 & 1.5 & 1.25 \\
   \hline
   $\lambda'$ (nm) (QB) & 446.339 & 451.60 & 456.60 & 456.79 \\
   \hline
   $Q$ & 8047 & 161.28 & 87.8 & 29.2 \\
   \hline
   $\lambda'$ (nm) (laser) & 446.335 & 451.60 & 456.5 & 455.827 \\
   \hline
   $n''_l$ & -1.967$\times10^{-4}$ & -0.0055 & -0.0124 & -0.0378 \\
   \hline
   $\left<\sigma_d\right>$ (\%) & 0.05 & 3 & 8.4 & 14.5 \\
   \hline
 \end{tabular}
\caption{Wavelength $\lambda'$ and quality factor $Q$ of the QB states; lasing frequency $\lambda'$ and imaginary part of the refractive index $n''_l$ obtained for the threshold lasing modes; relative index difference $\left<\sigma_d\right>$ between QB states and TLM, for four index values $n'$ of the scatterers.}
\label{Table1}
\end{table}

The values of wavelengths and quality factors of the QB states, lasing frequencies of the corresponding threshold lasing modes and associated imaginary part of the refractive index are summarized in Table \ref{Table1}, together with the relative difference $\left<\sigma_d\right>$ as defined in Eq.\ref{error}.

\begin{figure}[h]
\centering\includegraphics[clip,width=.5\linewidth]{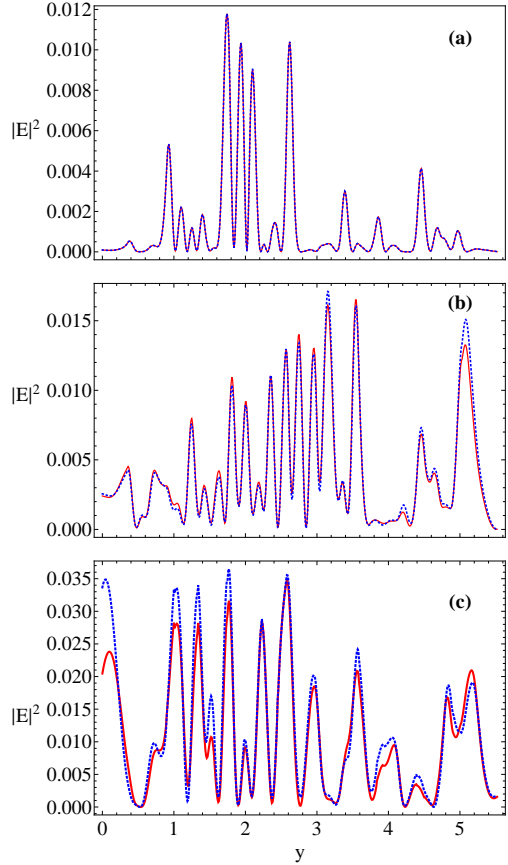}
\caption{The intensity $|E|^2$ of  QB state (blue dashed line) and lasing mode (red solid line) at $x=2.75$ for $n'_l=1.75$ (a), $n'=1.5$ (b) and $n'=1.25$ (c).}
\label{Fig7_Ara}
\end{figure}
In order to visualize the increasing difference between TLM and QB states, the cross-section of their spatial intensity profile at $x=2.75$ is plotted in Fig.~\ref{Fig7_Ara}. In Fig.~\ref{Fig7_Ara}(a) one cannot distinguish between the lasing mode and the QB state for $n'=1.75$, while for $n'=1.5$ (Fig.~\ref{Fig7_Ara}(b)) differences begin to emerge, becoming more pronounced for the case of $n'=1.25$ (Fig.~\ref{Fig7_Ara}(c)). This is seen also in the increase of the relative difference from 5\% to 14.5\%. Clearly, there is a systematic increase of the discrepancy between QB states and lasing modes when index contrast and scattering decrease and leakage increases. For very low scattering $n'=1.05$, we could not find the QB state corresponding to the TLM. Although we may have missed a pole in the complex plane, this raises however a serious question on the validity of the comparison of the threshold laser mode with QB states when weakly scattering systems are considered. In the next section, we will argue that, in principle, QB states cannot be the support of the TLM. Section \ref{sec:multimode} will introduce a different class of states, which offer a valid basis on which the TLMs can be described.

\section{Threshold lasing states vs. passive cavity resonances}
\label{sec:TLMvsQB}

Semiclassical laser theory treats classical electromagnetic fields coupled to quantized matter and yields the thresholds, frequencies and electric fields of the lasing modes, but not their linewidths or noise properties.  In order to treat the spatial dependence of lasing modes one must go beyond rate equation descriptions and use the coupled non-linear Maxwell-Bloch (MB) equations for light coupled to homogeneously broadened two-level ``atoms" or multilevel generalizations thereof. These equations will be presented in section \ref{sec:multimode} below. While the MB description has been used since the inception of laser theory \cite{Lamb,Haken}, in almost all cases simplifications to these equations were made, most notably a neglect of the openness of the laser cavity.  As random lasers are strongly open systems, it is necessary to treat this aspect of the problem correctly to obtain a good description of them.

Historically a first breakthrough in describing Fabry-Perot type lasers with open sides was the Fox-Li method \cite{FoxLi1960}, which is an integral equation method of finding the passive cavity resonances of such a structure. It is widely assumed and stated that these resonances or quasi-bound (QB) states are the correct electromagnetic modes of a laser, at least at threshold. Often the non-linear laser equations are studied with hermitian cavity modes with phenomenological damping constants representing the cavity outcoupling loss obtained, e.g. from a Fox-Li calculation.  It is worth noting that there are two kinds of cavity loss that occur in lasers; there is the outcoupling loss just mentioned and also the internal absorption of the cavity which can be taken into account via the imaginary part of the passive cavity index of refraction. These are very different processes as the former describes the usable coherent light energy emitted from the laser and the latter simply energy lost, usually as heat, in the laser cavity.

The QB states of an arbitrary passive cavity described by a linear dielectric function $\eps_c (\bx, \omega)$ can be rigorously defined in terms of an electromagnetic scattering matrix $S$ for the cavity.  This matrix relates incoming waves at wavevector $k$ (frequency $\omega = ck$) to outgoing waves in all of the asymptotic scattering channels and can be calculated from the wave equation.  Note that while we speak of the frequency of the incoming wave in fact the S-matrix is a {\it time-independent quantity} depending on the wavevector $k$. This is the wavevector outside the cavity; in random lasers we will be interested in spatially varying dielectric functions so that in the ``cavity" there is no single wavevector of the field. For any laser, including the random laser, the cavity can be defined as simply the surface of ``last scattering" beyond which no backscattering occurs. The QB states are then the eigenvectors of the passive cavity S-matrix with eigenvalue equal to infinity; i. e. one has outgoing waves with no incoming waves. Because this boundary condition is incompatible with current conservation these eigenvectors have complex wavevector, $ \tkmu$; these complex frequencies are the poles of the S-matrix and their imaginary parts must always be negative to satisfy causality conditions.  There are normally a countably infinite set of such QB states. Due to their complex wavevector, asymptotically the QB states vary as $r^{-(d-1)/2}\exp (+| Im[\tkmu]| r)$ and diverge at infinity, so they are not normalizable solutions of the time-independent wave equation.  Therefore we see that QB states cannot represent the lasing modes of the cavity, even at threshold, as the lasing modes have real frequency and wavevector outside the cavity with conserved photon flux.

\begin{figure}[h]
\centering\includegraphics[clip,width=.8\linewidth]{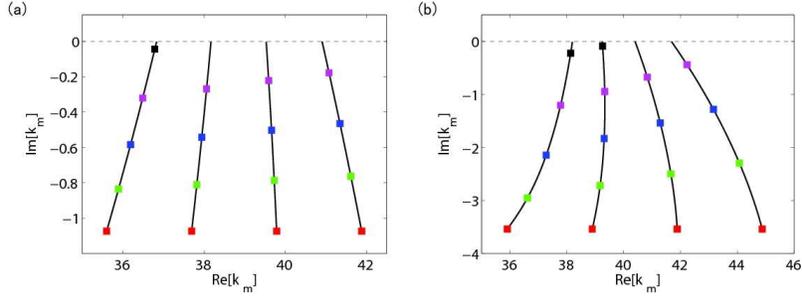}
\caption{Shift of the poles of the S-matrix in the complex plane onto the real axis to form threshold lasing modes when the imaginary part of the dielectric
function $\epsilon \equiv \epsilon_c + \epsilon_g$ varies for a simple 1d edge-emitting cavity laser \cite{Li08}.  The cavity is a region of length $L$ and uniform index $n_c=1.5 (a), 1.05 (b) (\eps_c = 2.25, 1.0025)$ terminated in vacuum at both ends.  The calculations are based on the MB model discussed in section \ref{sec:multimode}, with parameters $k_aL=39$ and $\g=2$. (a) $n_c=1.5$, the squares of different colors represent $\mathrm{Im}[\epsilon_g]=0,-0.032,-0.064,-0.096,-0.128$  (b) $n_c=1.05$,the squares of different colors represent $\mathrm{Im}[\epsilon_g]=0,-0.04,-0.08,-0.12,-0.16$, note the increase in the frequency shift in the complex plane for the leakier cavity. The center of the gain curve is at $kL = 39$ which determines the visible line-pulling effect.} \label{fig:SmatrixPoles}
\end{figure}

When gain is added to the cavity the effect is to add another contribution to the dielectric function $\eps_g (\bx,\omega)$ which in general has a real and imaginary part.  The imaginary part of $\eps_g$ has amplifying sign when the gain medium is inverted and depends on the pump strength; it compensates the outcoupling loss as well as any cavity loss from the cavity dielectric function $\eps_c$.  The specific form of this function for the MB model will be given in section \ref{sec:multimode} below. The threshold lasing modes (TLMs) are the solutions of the wave equation with $\eps_{total}(\bx) = \eps_c(\bx) + \eps_g(\bx)$ with only outgoing waves of {\it real} wavevector $k_\mu$ (we neglect henceforth for simplicity the frequency-dependence of $\eps_c(\bx)$).  The $k_\mu$ are the wavevectors of the TLMs with real lasing frequencies $\Omega_\mu = c k_\mu$.  These lasing wavevectors are clearly different from the complex $\tkmu$; moreover they are not equal to $Re[\tkmu]$ as often supposed. This can be seen by the following continuity argument.  Assume that $\eps_c(\bx)$ is purely real for simplicity, so that the S-matrix is unitary and all of its poles are complex and lie in the negative half plane.  Turn on the pump, which we will call $D_0$ anticipating our later notation, so that the inversion rises steadily from zero, continuously increasing the amplifying part of $\eps_g$. The S-matrix is no longer unitary, and its poles move continuously ``upward" towards the real axis until each of them crosses the axis at a particular pump value, $D_0$ (see Fig. \ref{fig:SmatrixPoles}); the place where each pole crosses is the real lasing frequency $k_\mu$ for that particular TLM. Note that the poles do not move vertically to reach the real axis but always have some shift of the lasing frequency from the passive cavity frequency, mainly due to line-pulling towards the gain center. As the Q-value of the cavity increases, the distance the poles need to move to reach the real axis decreases so that the frequency shift from $Re[\tkmu]$ can become very small and the conventional picture becomes more correct.  In general the poles of the S-matrix are conserved quantities even in the presence of loss, so that the TLMs are in one to one correspondence with the QB states and thus are countably infinite, but for any cavity the pole which reaches the real axis first (i.e. at lowest pump $D_0$) is the actual first lasing mode.  At higher pump values the non-linear effects of saturation and mode competition will affect the behavior; so only the lowest threshold TLM describes an observable lasing mode for fixed pumping conditions, the {\it first lasing mode} at threshold.  Which pole gets there first depends not only on the Q of the passive cavity resonance before gain is added, but also on the parameters of $\eps_g (\bx)$ which include the atomic transition frequency, the gain linewidth and the pump conditions as will be discussed below.

\section{Self-consistent time-independent approach to random lasing}
\label{sec:multimode}

In section \ref{sec:TLMvsQB} we gave a general argument based on the scattering matrix with the addition of gain to show that in general the QB states (passive cavity resonances) are never exactly the same as the threshold lasing modes (TLMs), even inside the cavity.  However the same argument indicated that inside a high-Q cavity the two sets of functions become very similar since the poles of the S-matrix are very close to the real axis and only a small amount of gain is required to move them to the real axis, which maps QB states onto TLMs.  For localized states in the center of the sample the Q-values should be exponentially large and, as found numerically, QBs and TLMs should be indistinguishable (again, inside the cavity, outside the QB states have an unphysical growth).  As already noted, the set of TLMs only defines threshold modes, as soon as the first TLM has turned on it will alter the gain medium for the other potential modes through spatial hole-burning and a non-linear approach needs to be considered. Very recently such an approach has been developed which has the major advantage of being time-independent and partially analytic, providing both ease of computation and greater physical insight.  The approach, due to T\"ureci-Stone-Ge, is known as Ab Initio Self-consistent Laser Theory (AISC laser theory) \cite{Tureci06,Li08,Tureci09}.
It finds the
stationary solutions of the MB semiclassical lasing equations in the multimode regime, for cavities of arbitrary complexity and openness, and to infinite order in the non-linear interactions.  As such it is ideal for treating diffusive or quasi-ballistic random lasers which are extremely open and typically highly multimode even slightly above threshold.  In this section we present the basic ideas with emphasis on threshold lasing modes, which is the focus of this review. The non-linear theory has been reviewed in some detail elsewhere \cite{Tureci09}, and we just present a brief introduction to it here.

\subsection{Maxwell-Bloch threshold lasing modes}

The MB semiclassical laser equations describe a gain  medium of identical two-level ``atoms'' with energy level spacing $\hbar \omega_a = \hbar c k_a$ and relaxation rate $\gp$, being pumped by an external energy source, $D_0$ (which can vary in space), contained in a cavity which can be described by a linear dielectric function, $\eps_c (\bx)$. This leads to a population inversion of the atoms, $D(\bx,t)$ which in the presence of an electric field creates a non-linear polarization of the atomic medium, $P(\bx,t)$, which itself is coupled non-linearly to the inversion through the electric field, $E(\bx,t)$.  The electric field and the non-linear polarization are related linearly through Maxwell's wave equation, although above the first lasing threshold the polarization is implicitly a non-linear function of the electric field.  The induced polarization also relaxes at a rate $\g$ which is typically much greater than the rate $\gp$ at which the inversion relaxes, and this is a key assumption in our treatment of the non-linear regime, but will not be needed in the initial discussion of TLMs.

The resulting system of non-linear coupled partial differential equations for the three fields $E(\bx,t),P(\bx,t),D(\bx,t)$ are ($c=1$):
\bea
\ddot{E}^+  &=& \frac{1}{\eps_c (\bx)}\nabla^2 {E}^+ - \frac{4\pi}{\eps_c (\bx)} \ddot{P}^+
\label{eqMB0}\\
\dot{P}^+ &=& -(i\om_a + \g) \Pp + \frac{g^2}{i\hbar} \Ep D  \label{eqMB1}\\
\dot{D} &=& \gp\left( D_0-D \right) - \frac{2}{i\hbar}\left( \Ep (\Pp)^* -
\Pp (\Ep)^* \right)\,. \label{eqMB2}
\eea
Here $g$ is the dipole matrix element of the atoms and the units for the pump are chosen so that $D_0$ is equal to the time-independent inversion of the atomic system in the absence of an electric field.  This pump can be non-uniform: $D_0 = D_0(\bx)$ based on the experimental pump conditions, but we will not discuss that case here. The electric field, polarization and inversion are real functions ($E,P$ are vector functions in general, but we assume a geometry where they can be treated as scalars).  In writing the equations above we have written these fields in the usual manner in terms of their  positive and negative frequency components, $E =E^+ + E^-$, $P = P^+ + P^-$, and then made the rotating wave approximation (RWA) in which the coupling of negative to positive components is neglected.  There is no advantage in our treatment to making the standard slowly-varying envelope approximation and we do not make it.

\subsection{Self-consistent steady-state lasing equations}

The starting point of our formulation is to assume that there exists a steady state multiperiodic solution of equations (\ref{eqMB0})-(\ref{eqMB2}) above, i.\,e., we try a solution of the form:
\be
\Ep (\bx, t) = \sum_{\mu=1}^N \psimu e^{-i k_\mu t},  \;\;\;\;\Pp (\bx, t) = \sum_{\mu=1}^N P_\mu (\bx) e^{-i k_\mu t}\,.
\label{eqEPansatz0}
\ee
Having taken  $c=1$ we do not distinguish between frequency and wavevector.  The functions $\psimu$ are the unknown lasing modes and the real numbers $k_{\mu}$ are the unknown lasing frequencies; these functions and frequencies are not assumed to have any simple relationship to the QB states of the passive cavity and will be determined self-consistently.  As the pump increases from zero the number of terms in the sum will vary, $N=0,1,2,\ldots$ at a series of thresholds each new mode will appear. The general non-linear theory is based on a self-consistent equation which determines how many modes there are at a given pump, and solves for these modes and their frequencies. However in this section we will discuss threshold lasing modes (TLMs) and so we need only consider one term in the sum. Furthermore, at the first threshold the electric field is negligibly small and so the inversion is equal to the external pump profile, assumed uniform in space, $D(\bx,t) = D_0 $.   Assuming single-mode lasing the equation for the polarization becomes:
\be
P_\mu (\bx) = \frac{-iD_0 g^2 \psimu }{\hbar (\g - i(k_\mu - k_a))}
\ee
Having found $P_\mu (\bx)$ in terms of $\psimu, D_0,$ we substitute this result into the right hand side of Maxwell's equation along with $\psimu$ for the electric field on the left hand side.  The result is:
\be
[\nabla^2 +  \eps_c(\bx)k_\mu^2]\psimu = \frac{iD_0 4 \pi g^2 k_\mu^2 \psimu }{\hbar (\g - i(k_\mu - k_a))},
\ee
which can be written in the form:
\be
[\nabla^2 +  (\eps_c(\bx) + \eps_g (\bx))k_\mu^2]\psimu = 0,
\label{eqdiffTLM}
\ee
where $\eps_g (\bx)$ is the dielectric function of the gain medium, which only varies in space if the external pump or the gain atoms are non-uniform.  Defining convenient units of pump $D_{0c} = \hbar \g/4 \pi k_a^2 g^2$ and replacing $D_0 \Rightarrow D_0/D_{0c}$, we find that
\be
\eps_g(\bx) =  \frac{D_0}{k_a^2}[\frac{\g(k_\mu - k_a)}{\g^2 + (k_\mu - k_a)^2} + \frac{-i \g^2}{\g^2 + (k_\mu - k_a)^2}].
\label{eqepsg}
\ee

Equation (\ref{eqdiffTLM}) is to be solved with the boundary condition that at infinity one has only an outgoing wave at frequency $k_\mu$, i.\,e., $\nabla_r \psimu = +ik_\mu \psimu$ when $r \to \infty$. In general this equation with this boundary condition cannot be solved for arbitrary choice of the lasing frequency $k_\mu$ and for arbitrary values of the  pump $D_0$; it is necessary to vary $k_\mu$ and the pump strength $D_0$ to find the countably infinite set of values $(k_\mu,D_0^{(\mu)})$ at which a solution exists.  This variation is equivalent to the pulling of the S-matrix poles onto the real axis discussed in section \ref{sec:TLMvsQB} above; $D_0^{(\mu)}$ defines the threshold pump, {\it for that pole}, and $k_\mu$ the point at which it crosses the real axis.  As noted, while all of these solutions can be classified as TLMs, only the solution with the lowest value of $D_0^{(\mu)}$ will actually be a physical lasing state, as higher lasing modes are altered by non-linear modal interactions.

Equation (\ref{eqdiffTLM}) shows that the TLMs are the solutions of the original Maxwell equation with the addition of a complex, pump and frequency dependent dielectric function which is uniform in space (for the assumed uniform pumping).  The imaginary  and real parts of the gain dielectric function have the familiar symmetric and anti-symmetric two-level resonance form respectively. The dependence on the atomic frequency $k_a$ encodes the usual atomic line-pulling effect.  In the limit of a very broad gain curve ($\g \to \infty$) the line-pulling effects can be neglected and we find the simple result
\be
\eps_g \to -i D_0/k_a^2 ,
\ee
i.e. a constant imaginary (amplifying) part of $\eps_g$ proportional to the pump strength.  Such linear gain models have been studied before, although typically with a constant imaginary part of the index of refraction instead of a constant imaginary part of the dielectric function.  Our results show that in order to reproduce the TLMs of the MB equations one needs to take
\be
n(\bx) = \sqrt{\eps_c(\bx) +\eps_g (D_0, k_\mu-k_a, \g) }
\ee
so that the pump changes both the real and imaginary parts of the index of refraction.

\subsection{Solution for TLMs and CF states}

\begin{figure}[h]
\centering\includegraphics[clip,width=0.5\linewidth]{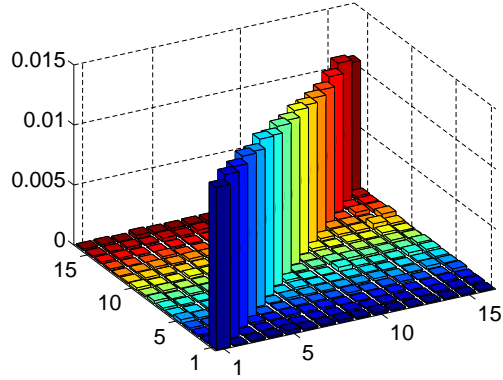}
\caption{Typical values of the threshold matrix elements ${{\bf \cal T}}^{(0)}$ in a two-dimensional random laser schematized in the inset Fig. \ref{fig:CF_frequencies} using sixteen CF states. The off-diagonal elements are one to two orders of magnitude smaller than the diagonal ones.}
\label{fig:diagonality}
\end{figure}

\begin{figure}[h]
\centering\includegraphics[clip,width=.8\linewidth]{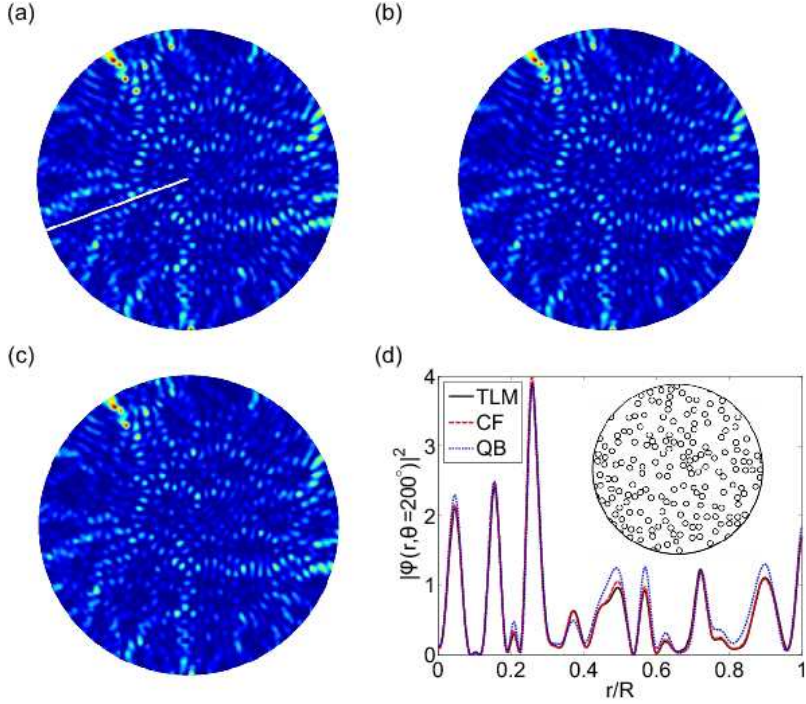}
\caption{(a) False color plot of one TLM in a 2D random laser modeled as an aggregate of sub-wavelength particles of index of refraction $n = 1.2$ and radius $r=R/30$ against a background index $n =1$ imbedded in a uniform disk of gain material of radius R(see inset, Panel (d)). The frequency of the lasing mode is $kR=59.9432$, which is pulled from the real part of the dominating CF state $k_mR = 59.8766 - 0.8593i$ (b) towards the transition frequency $k_aR=60$. The spatial profile of the TLM and CF state agree very well, whereas the corresponding QB state $\tilde{k}_mR = 59.8602 - 0.8660i $ (c) differs from that of the TLM and the CF state noticeably, as can be seen in Panel (d) where we plot the internal intensity along the $\theta=200^{\circ}$ direction (white line in panel (a)).}
\label{fig:QBCFTLM}
\end{figure}
\begin{figure}[h]
\centering\includegraphics[clip,width=.8\linewidth]{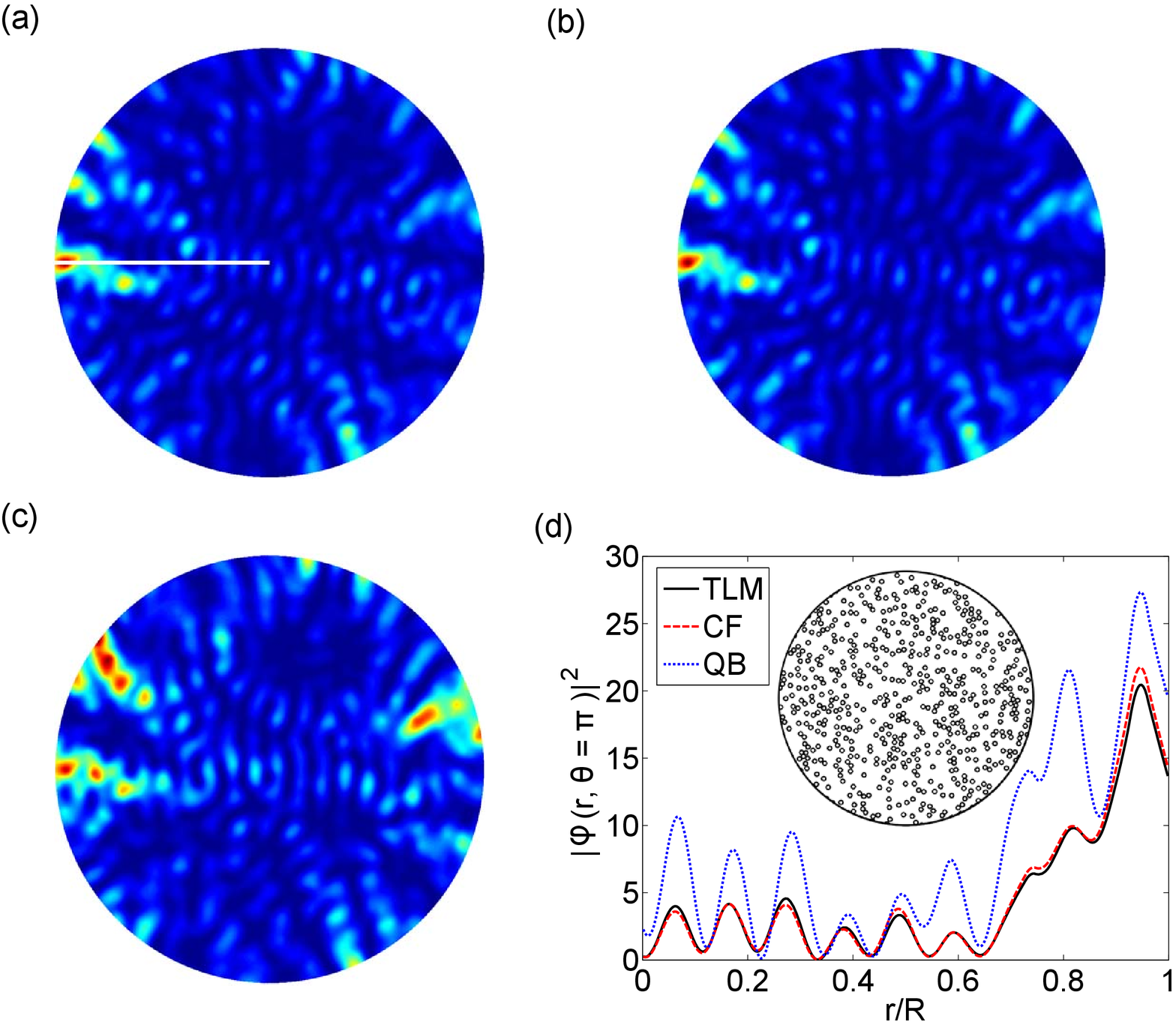}
\caption{(a) False color plot of one TLM in a 2D random laser similar to that in the Fig. \ref{fig:QBCFTLM} but with particles of radius $r=R/60$ corresponding to weaker scattering (see inset, Panel (d)). The frequency of the lasing mode is $kR=29.9959$, which is very close to the CF state $k_mR = 30.0058 - 1.3219i$ (b) but shifted from the corresponding QB state $\tilde{k}_mR = 29.8813 -1.3790i$ (c). In Panel (d) we show the internal intensity of the three states in the $\theta=\pi$ direction (white line in panel (a)); due to weaker scattering the QB state now differs substantially from the CF and TLM, which still agree quite well with each other}
\label{fig:QBCFTLM_small}
\end{figure}

The differential equation (\ref{eqdiffTLM}) is self-consistent in the sense that the boundary conditions depend on the eigenvalue $k_\mu$ that one is solving for and so some form of non-linear search is required.  The required search turns out to be much more convenient if one writes an equivalent integral form of the equation transforming it into a self-consistent eigenvalue problem.  For this purpose we rewrite it in the form
\be
[ \eps_c(\bx)^{-1} \nabla^2 + k_\mu^2]\psimu = \frac{-\eps_g k_\mu^2}{\eps_c(\bx)}\psimu,
\ee
and then, treating the right hand side as a source, invert the equation with the appropriate Green function to obtain:
\begin{equation}
\Psi_\mu (\bx) =  \frac{iD_0 \g}{\g - i (k_\mu -
k_a)}\frac{k_\mu^2}{k_a^2} \int_{\cal D} d {\bxp} \frac{
G(\bx,\bxp;k_\mu)  \Psi_\mu (\bxp)}{\eps_c(\bxp)}.
\label{eqTLM}
\end{equation}
Here the integral is over the gain region which we will assume coincides with the cavity region ${\cal D}$.  The appropriate Green function satisfies
\be
[ \eps_c (\bx)^{-1}  \nabla^2  +  k^2 ]\,
G(\bx,\bx'| k) = \delta^d(\bx-\bx'),
\ee
and is non-hermitian due to the outgoing wave boundary conditions: $\nabla_r G(\bx,\bx'| k)|_{r \to \infty} = \nabla_{r \prime} G(\bx,\bx'| k)|_{r \prime \to \infty} = ikG(\bx,\bx'| k)$, where $ \nabla_r$ is the radial derivative.  It has the spectral representation:
\begin{equation}
G(\bx,\bxp |k) =  \sum_m
\frac{\varphi_{m}(\bx,k) \bar{\varphi}^*_{m}(\bxp,k)}{(k^2 - k_m^2)}.
\label{eqspecrep2}
\end{equation}
We refer to the functions $\varphi_{m}(\bx,k)$  in (\ref{eqspecrep2}) as the constant-flux (CF) states.  They satisfy
\be
[ \eps_c(\bx)^{-1}  \nabla^2 + k_m ^2]\, \varphi_m(\bx,k) =0
\label{eqcf}
\ee
with the corresponding non-hermitian boundary condition of purely outgoing spherical waves of fixed frequency $k$ (eventually set equal to the lasing frequency) at infinity.  Their dual (biorthogonal) partners $\bar{\varphi}_{m}(\bxp,k)$ satisfy the complex conjugate differential equation with purely incoming wave boundary conditions.  These dual sets satisfy the biorthogonality relation:
\be
\int_{
\cal D} d\bx \;\varphi_m(\bx,k) \bar{\varphi}^*_{n}(\bx,k) = \delta_{mn} \label{eq:biorth}
\ee
with appropriate normalization.

The CF states satisfy the standard wave equation Eq.~(\ref{eqcf}), but with the non-hermitian boundary condition already mentioned; hence their eigenvalues $k_m^2$ are complex, with (it can be shown) negative imaginary part, corresponding to amplification within the cavity.  However outside the cavity, by construction, they have the real wavevector $k_\mu$ and a conserved photon flux.  They are a complete basis set for each lasing frequency $k_\mu$ and hence they are a natural choice to represent the TLMs, as well as the lasing modes above threshold.  Hence we make the expansion
\begin{equation}
\Psi_\mu (\bx) = \sum_{m=1}^\infty a_m^\mu \varphi_m^{\mu} (\bx)\,
\label{eqmpansatz}
\end{equation}
Substituting this expansion into Eq.~(\ref{eqTLM}), using biorthogonality, and truncating the expansion at $N$ terms leads to the eigenvalue problem:
\be
a^\mu_m = D_0 \Lambda_m (k_\mu)  \int_{{\cal D}}d {\bxp}
\frac{ \bar{\varphi}_m^{\mu*} (\bxp) \sum_p^N a^\mu_p
\varphi^\mu_p(\bxp)
}{\eps_c(\bxp)}\ \equiv D_0 \sum_{p}^N {\cal T}^{(0)}_{mp}a^\mu_p,
\label{eqamTLM}
\ee
where $\Lambda_m (k) \equiv i \g (k^2/k_a^2)/ [ (\g - i (k - k_a)) (k^2 - k_m^2(k)) ]$.

One sees that the TLMs in the CF basis are determined by the condition that an eigenvalue of the matrix   $D_0 {{\bf \cal T}}^{(0)}(k_\mu)$ is equal to unity.  Since the
matrix ${{\bf \cal T}}^{(0)}(k_\mu)$ is independent of $D_0$ it is natural to focus on this object, which we call the {\it threshold matrix}.  It is a complex matrix with no special symmetries, implying that its eigenvalues $\lambda_\mu$ are all complex for a general value of $k_\mu$.  If the real control parameter $D_0$ (the pump) is set equal to $1/|\lambda_\mu |$ then the matrix $D_0 {{\bf \cal T}}^{(0)}(k_\mu)$ will have an eigenvalue of modulus unity, but not a real eigenvalue equal to unity as required and no solution for the TLMs exists for this choice of $k_\mu$.  It is the phase condition, that $\lambda_\mu(k_\mu)$ must be real that determines the allowed lasing frequencies.  In practice one orders the $\lambda_\mu$ in decreasing modulus based on an initial approximation to the lasing frequency, $k_\mu$, and then tunes $k_\mu$ slowly until each eigenvalue flows through the real axis (which  is guaranteed by the dominant k-dependence contained in the factor $\Lambda_m (k)$).  Normally the eigenvalues do not switch order during this flow and the largest eigenvalue $\lambda_\mu$ will determine the lowest threshold TLM, with threshold $D_0^{(\mu)} = 1/ \lambda_\mu (k_\mu)$, where $k_\mu$ is the frequency which makes the largest eigenvalue  ${{\bf \cal T}}^{(0)}(k_\mu)$ real. The eigenvector corresponding to $\lambda_\mu$ gives the coefficients for the CF expansion of the TLM of the first mode $\psimu$.  TLMs with higher thresholds can be found by imposing the reality condition on smaller eigenvalues of  ${{\bf \cal T}}^{(0)}(k_\mu)$. This approach has been described in detail elsewhere \cite{Tureci08,Tureci09}, and provides a much more efficient method for finding TLMs than solving the self-consistent differential equation, (\ref{eqdiffTLM}).

We immediately see from Eqs. (\ref{eq:biorth}) and (\ref{eqamTLM}) that for an arbitrarily shaped cavity of uniform dielectric constant $\eps_c$ the matrix ${\bf {\cal T}}^{(0)}(k_\mu)$ is diagonal due to biorthogonality of the CF states.  Thus each TLM is a single CF state, corresponding to one of the $k_\mu$ which satisfies the reality condition.  In this case the expansion of $\psimu$ consists of just one term and the threshold lasing equation is equivalent to the Eq.~\ref{eqcf} with appropriate relabeling. When $\eps_c$ varies in space, as for random lasers, the threshold matrix is not diagonal and there can in principle be many CF states contributing to one TLM.  However since $\varphi_m (\bx), \bar{\varphi}_p (\bx)$ are uncorrelated fluctuating functions of space, it turns out that the threshold matrix in RLs is approximately diagonal and the threshold modes are dominated by one, pseudo-random CF state determined by solving Eq.~\ref{eqcf} for the appropriate random dielectric function $\eps_c (\bx)$.  This is shown in Fig. \ref{fig:diagonality} below. In summary, the theory leading to the threshold equation (\ref{eqamTLM}) gives an efficient time-independent method for finding the TLMs of random lasers in any disorder regime. In general these TLMs are very close to a single CF state determined by the Eq.~\ref{eqcf} at the lasing frequency $k_\mu$.With this new method TLMs of random lasers can be found for complex two and even three-dimensional geometries. In Figs. \ref{fig:QBCFTLM} and \ref{fig:QBCFTLM_small} we compare TLMs, CF states and QB states for the two-dimensional random laser model used in ref. \cite{Tureci08}, illustrating the agreement of TLMs with CF states even for weak scattering, while a significant deviation from the closest QB state is found.

This AISC laser theory is well-suited to describe not just TLMs but to find the true multimode lasing spectrum of RLs {\it above threshold}. This will not be treated in detail here, but in the next section we briefly explain the basic approach in the non-linear theory and show one representative result.

\subsection{Non-linear AISC laser theory}

The key to generalizing this theory to the multi-mode non-linear regime is to return to the fundamental MB equations and go beyond the assumption that the inversion $D (\bx, t)$ is equal to the constant threshold pump $D_0$.  Once lasing modes have turned on their spatially varying electric fields cause varying degrees of stimulated emission from the gain atoms and hence tends to reduce the inversion $D$ from the pump value $D_0$ in a manner which varies in space and in principle in time.  However it has been shown that if $\g \gg \gp$, then the time-dependence of the inversion is weak and although D is varies in space, it is a good approximation to take $D(\bx, t) = D(\bx)$.  This stationary inversion approximation (SIA) has been used in laser theory for many years, going back to Haken \cite{Haken}, but has not been incorporated into an ab initio method such as AISC laser theory. We will not review the details of the derivation of the non-linear multimode theory of T\"ureci-Stone-Ge, which have been given elsewhere \cite{Tureci06,Tureci09}.  Instead we just state that the net effect of the non-linear interactions within the SIA is just to replace the uniform inversion as follows:
\be
D_0 \to \frac{D_0}{1 +   \sum_{\nu} \Gamma (k_\nu) |\Psi_\nu(\bx)|^2)}
\label{eqsub}
\ee
in all of the equations of the theory of the TLMs.  Here $\nu$ labels all above threshold modes and $\Gamma (k_\nu)$ is a Lorentzian centered at the lasing frequency of mode $\nu$ with width $\g$.  If we make this substitution into the Eq.~(\ref{eqTLM}) we arrive at the fundamental integral equation of AISC laser theory:
\begin{equation}
\Psi_\mu (\bx) =  \frac{i D_0\g}{\g - i (k_\mu -
k_a)}\frac{k_\mu^2}{k_a^2} \int_{\cal D} d {\bxp}
\frac{ G(\bx,\bxp;k_\mu)  \Psi_\mu (\bxp)}{\eps_c (\bxp) (1 + \sum_\nu
\Gamma_\nu |\Psi_\nu (\bxp)|^2)}\, . \label{eqTSG}
\end{equation}

\begin{figure}[h]
\centering\includegraphics[clip,width=.9\linewidth]{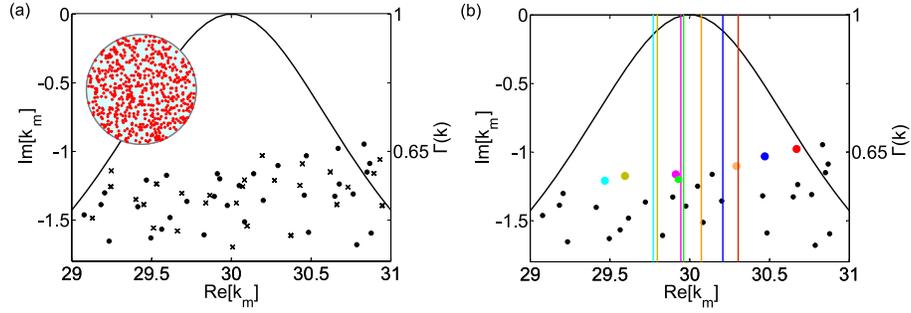}
\caption{(a) CF (dots) and QB (crosses) frequencies in a 2D random laser modeled as an aggregate of sub-wavelength particles of index of refraction $n = 1.2$ against a background index $n =1$ imbedded in a uniform disk of gain material (see inset). The two sets of complex frequencies are statistically similar but differ substantially. The solid curve shows the gain curve $\Gamma(k)$ with $\gamma_\perp=1$. (b) Lasing frequencies of the same random system well above threshold (coloured lines). Coloured circles denote the CF state dominating the correspondingly colored modes at threshold.}
\label{fig:CF_frequencies}
\end{figure}

Note that this equation shows that each lasing modes interacts with itself (saturation) and all other lasing modes (mode competition) via the ``hole-burning" denominator of Eq.~ (\ref{eqsub}).  This set of coupled non-linear equations is still conveniently solved in the basis of CF states for each modal frequency $k_\mu$, as for the TLMs; the details have been given elsewhere \cite{Tureci08,Tureci09}.

The first results of the AISC laser theory for the modal properties of multimode random lasers in weak-scattering two-dimensional media were given in \cite{Tureci08}.  We will not present a full picture of these results here, but just show some properties of the RL lasing frequencies in Fig. \ref{fig:CF_frequencies} below. The model is explained in the figure caption (see inset).  The complex CF and QB frequencies are shown to be distinct and the lasing frequencies are subject to very strong line-pulling effects.

The new tool of AISC laser theory allows one to study random lasers with full non-linear interactions in 2D and even in 3D.  The elimination of time-dependence in this theory makes larger and more complex cavities computationally tractable. The theory also provides a new language based on CF states to describe the lasing modes. Now detailed statistical studies as well as comparisons to statistical models based on random matrix theory, disordered media theory and wave chaos theory are needed.  Such studies are in progress.

\section{Conclusion}

A decade of theoretical study of random lasers has clarified the nature of the lasing modes in disordered systems with multiple scattering and gain.  Most importantly it has been established that high-Q passive cavity modes such as those created by Anderson Localization or by rare fluctuations of various kinds are not necessary in order to have self-organized laser oscillation at a frequency distinct from the atomic transition frequency (gain center). In addition this study has emphasized a point of general importance in laser theory, that threshold lasing modes are not identical to the quasi-bound states (resonances) of the passive cavity. This point is demonstrated by a number of numerical calculations presented above and also can be understood from the realization that the QB states are eigenvectors of the unitary S-matrix of the cavity without gain, but at complex frequency, whereas the threshold lasing modes are eigenvectors of the non-unitary S-matrix of the cavity {\it with gain} and with real frequency.  The difference between these eigenvectors (within the cavity), which is large in the weak scattering limit, becomes small in the diffusive regime as the Q of the cavity increases and is negligible, e.g. for Anderson localized modes and for high-Q modes of conventional cavities. The new basis set of {\it constant flux} states provides a better approximation for finding the threshold lasing modes of random lasers and coincides with the exact lasing modes of uniform index cavities. Further statistical and analytical study is necessary to characterize the properties of random lasers in the different regimes, weak scattering, diffusive and localized, and to understand the effects of non-linear interactions.

\section{Acknowledgments}
P. Sebbah thanks the French National Research Agency which supports this work under grant ANR-08-BLAN-0302-01, the PACA region, and the CG06. A. A. Asatryan acknowledges the support from the Australian Research Council under its Centres of Excellence and Discovery Grants programs.  A. D. Stone acknowledges support from the National Science Foundation under grant DMR-0908437. H. E. T\"ureci acknowledges support from the Swiss NSF under Grant No. PP00P2-123519/1.

\appendix

\section{The multipole method}\label{apA}
This appendix details the principle of the multipole method as used in this paper and its implementation.  Although we describe here the method for two-dimensional systems, it can be also applied to three-dimensional structures.

\begin{figure}[h]
\centering\includegraphics[clip,width=.4\linewidth]{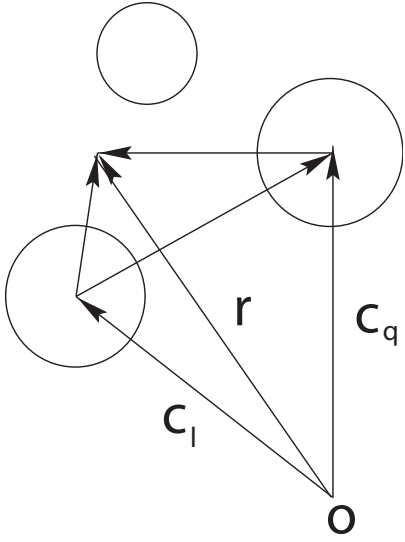}
\caption{The geometry and local co-ordinate systems.}
\label{Fig1_Ara}
\end{figure}

We consider a random collection of $N_c$ non-overlapping cylinders with arbitrary complex dielectric constant $\varepsilon_l=\varepsilon_l{'}+i \varepsilon_l{''}=n_l^2$ and arbitrary radii $a_l$ located in a uniform medium with complex dielectric constant $\varepsilon_b=\varepsilon_b{'}+i \varepsilon_b{''}=n_b^2$ (Fig. \ref{Fig1_Ara}), where $n_l=n_l'+in_l^{''}$ and $n_b=n_b'+in_b^{''}$ are the refractive indices of the cylinders and the background. The complex dielectric permittivities of the cylinders and the background can be arbitrary and may be frequency dependent.

In two dimensions, the solution of the electromagnetic field problem decouples into two fundamental polarizations, in each of which the field may be characterized by a single field component: $V({\bf r})=E_z$ (for TM polarization) and  $V({\bf r})=H_z$ (for TE polarization). In the co-ordinate system that is used, the $z$ axis is aligned with  the cylinder axes.

The field component $V$ satisfies the Helmholtz equation
\begin{equation}
\nabla^2 V({\bf r}) + k^2 n^2({\bf r})
 V({\bf r})=0.
\label{2}
\end{equation}
For TM polarization, both $ V({\bf r}) $ and its normal derivative ${\bf  \nu }\cdot \nabla V({\bf r}) $ are continuous across all boundaries, while for TE polarization the corresponding boundary conditions are the continuity of $V({\bf r}) $ and its weighted normal derivative $ {\bf \nu }\cdot \nabla V/n^2({\bf r}) $. Here, $n({\bf r})$ denotes the refractive index of the relative medium and ${\bf \nu} $ is an unit outward normal vector.

In the vicinity of the $l^{\rm th}$ cylinder, we may represent the exterior field in the background medium (refractive index $n_b$) in local coordinates $ {\bf r}_l =(r_l, \theta_l)={\bf r}-{\bf c}_l $ where ${\bf c}_l$ represents the center of the cylinder and write
\begin{equation}
V( {\bf r}) = \sum_{m=-\infty}^{\infty} \left[A^l_m J_m(k n_b r_l) +
B^l_m H^{(1)}_m(k n_b r_l)\right] e^{i m\theta_l}.
 \label{leo}
\end{equation}
This local expansion is valid only in an annulus extending from the surface of the cylinder $l$ to the surface of the nearest adjacent cylinder.

The global field expansion (also referred to as a Wijngaard expansion), which is valid everywhere in the background matrix, comprises only outgoing cylindrical harmonic terms:
\begin{equation}
V( {\bf r})   =  \sum_{q=1 }^{N_c} \sum_{m=-\infty}^{\infty}B^q_m
H^{(1)}_m(k|{\bf r}- {\bf c}_q|)e^{im \arg({\bf r}-{\bf c}_q)}.
\label{leoSM}
\end{equation}

Correspondingly, the field inside any cylinder $l$ is written in an interior expansion:
\begin{equation}
V({\bf r}) = \sum_{m=-\infty}^{\infty}\hspace{-2mm}
 C^l_m J_m(k n_l |{\bf r}-{\bf c}_l|) e^{im\arg({\bf r}-{\bf c}_l)}.
\label{lei}
\end{equation}

Then, applying Graf's addition theorem \cite{Asatryan03} to the terms on the right hand side of (\ref{leoSM}) (see Fig.~\ref{Fig1_Ara}),  we may express the global field expansion in terms of the local coordinate system for  the $l^{\rm th}$ cylinder. Equating this with the local expansion (\ref{leo}), we deduce the field identity (also known as the Rayleigh identity):
\begin{eqnarray}
 A^l_m & = & \sum_{q=1,q\neq l }^{N_c}\sum_{p=-\infty}^{\infty}
 H^{lq}_{mp}B^q_p,
  \label{leoSMa}
\end{eqnarray}
where
\begin{equation}
 H^{lq}_{mp} =
H^{(1)}_{m-p}(k c_{lq})e^{-i(m-p)\theta_{lq}}. \label{RSM1a}
\end{equation}
Here, $(c_{lq},\theta_{lq} )$ are the  polar coordinates of the vector ${\bf c}_{lq} =  {\bf c}_q - {\bf c}_l $, the position of cylinder $q$ relative to cylinder $l$.

This is the first connection between the standing wave ($\{A_m^l\}$) and outgoing ($\{B_m^l\}$) multipole coefficients, one which follows solely from the system geometry. Eq.~ (\ref{leoSMa}) indicates that the local field in the vicinity of cylinder $l$ is due to sources on all other cylinders ($q \ne l$), the contributions of which to the multipole term of order $m-p$ at cylinder $l$ are given by $H_{mp}^{lq}$.

The second relation between the $\{A_m^l\}$ and $\{B_m^l\}$ multipole coefficients is obtained from the field continuity equations (i.e., the boundary conditions) at the interface of cylinder $l$  and the local exterior  (\ref{leo}) and interior field (\ref{lei}) expansions. From these, we obtain:
\begin{eqnarray}
B^l_m & = & R^l_m A^l_m  \label{BC1}, \\
C^l_m & = &  T^l_m A^l_m  \label{BC2},
\end{eqnarray}
where the interface reflection and transmission coefficients, for both $E_z$ and $H_z$ polarization are given by
\begin{eqnarray}
R^l_m  & = & -\frac{\xi n_l J_m'(k n_l  a_l) J_m(k n_b a_l) -
n_bJ_m(k n_l a_l) J_m'(k n_b a_l)} {\xi n_l J_m'(k n_l  a_l)
H^{(1)}_m(k n_b a_l) - n_bJ_m(k n_l  a_l) H^{(1)'}_m(k n_b a_l)},
 \label{coef}\\
 {T^{l}_m}  &
= & -\frac{2i/(\pi k a_L)} {\xi n_l J_m'(n_l k a_l) H^{(1)}_m(kn_b
a_l) - n_bJ_m(k n_l a_l) H^{(1)'}_m(k n_b a_l)},
 \label{coefp}
\end{eqnarray}
in which $\xi$ = 1 for  TM polarization and $\xi=n_b^2({\bf r})/n^2_l({\bf r})$ for TE polarization.

To derive a simple closed form expression for the solution of the problem, we use partitioned matrix notation, introducing vectors ${\bf A}^l=[A^l_m]$ and ${\bf B}^l=[B^l_m]$   and expressing (\ref{leoSMa}) in the form
\begin{equation}
{\bf A }^l= \sum_q {\bf H}^{lq} {\bf  B }^q,
\end{equation}
where ${\bf A}^l$ and ${\bf B}^l$ denote vectors of multipole coefficients for cylinder $l$. The  matrix ${\bf H}$ is block partitioned according to ${\bf H}^{lq}= [H^{lq}_{mp}]$ for $l\ne q$  (\ref{RSM1a}), and ${\bf H}^{ll}= [{\bf 0}]$, each block of which is a matrix of Toeplitz form. Correspondingly, the matrix forms of Equations (\ref{BC1}) and (\ref{BC2}) are
\begin{eqnarray}
{\bf B}  & = & {\bf R} {\bf A}  \label{BC3}, \\
{\bf C}  & = &  {\bf T} {\bf A}  \label{BC4},
\end{eqnarray}
where ${\bf R}={\rm diag} \, {\bf R}^l$  is a block diagonal matrix of diagonal matrices ${\bf R}^l={\rm diag} \, R^l_m$, and with corresponding definitions applying for the transmission matrices.

Then, with the introduction of the  partitioned vectors ${\bf A} = [{\bf A}^l] $, ${\bf B} = [{\bf B}^l] $ and the partitioned matrix ${\bf H} = [{\bf H}^{lq}] $, we form the system of equations
\begin{eqnarray}
({\bf I} -{\bf R} {\bf H}){\bf B}= 0.
 \label{SM}
\end{eqnarray}
The problem has now been reduced to the solution of a generalized eigenvalue problem for the matrix equation (\ref{SM}). The nontrivial solutions of the secular equation (\ref{SM}) determines modes of the random system. Finding the nontrivial solutions of the linear system of equations (\ref{SM}) requires that the determinant of the system  matrix vanishes: (\ref{detSM})
\begin{eqnarray}
D&=& 0, \quad {\rm where~~} D = \det( {\bf S}^{-1}) \quad {\rm with }
\label{detSM}\\
{\bf S}^{-1}(\lambda) &=& ({\bf I} -{\bf RH}).
\end{eqnarray}
Equivalently, this problem may be recast as a search for the poles of the scattering matrix ${\bf S}(\lambda)$ (i.e., solutions of $\det{\bf S}^{-1}(\lambda)=0$).
Once the pole is located, the corresponding null vectors  $\bf B$ of (\ref{SM}) are the multipole coefficients of the scattered field which are used to calculate the QB state profiles exterior to the scatterers using (\ref{leoSM}). The field inside a cylinder is calculated according to the interior expansions (\ref{BC4}) and(\ref{lei}).
The TLM poles must be searched in the $(\lambda,\varepsilon_c)$ domain, given the pump changes not only the imaginary part of the refractive index but the real part as well (\ref{sec:TLMvsQB}).

The formal system (\ref{detSM}) is of infinite dimension and so must be truncated in order to generate a computational solution, the accuracy of which is governed by the number of retained multipole coefficients $N_m=2N_{\rm max}+1$, where $N_{\rm max}$ is the truncation order of the multipole series, i.e., only the terms corresponding to the cylindrical harmonics of order $n=-N_{\rm max}, ..., N_{\rm max}$ are retained.

\section{Finite Element Method} \label{femap}

We have also used the Finite Element Method (FEM) \cite{Jin93}, implemented in a commercial software (Comsol\texttrademark), to solve the wave equation (\ref{2}) and calculate the complex eigenvalues and eigenfunctions of the passive modes of the systems that were calculated by the multipole method. The method suitably applies for modeling passive or active modes in a cavity, which is surrounded by perfectly matched layers \cite{Pekel07} to simulate open boundaries. It is possible to obtain all the leaky modes, even the resonances characterized by a very small quality factor (as small as 5), in a reasonable computation time with a commercial PC, provided the size of the geometry is smaller than hundred times the wavelength. This is in contrast with the other methods described in this paper, which require much heavier computation.

\begin{figure}[h]
\centering\includegraphics[clip,width=.9\linewidth]{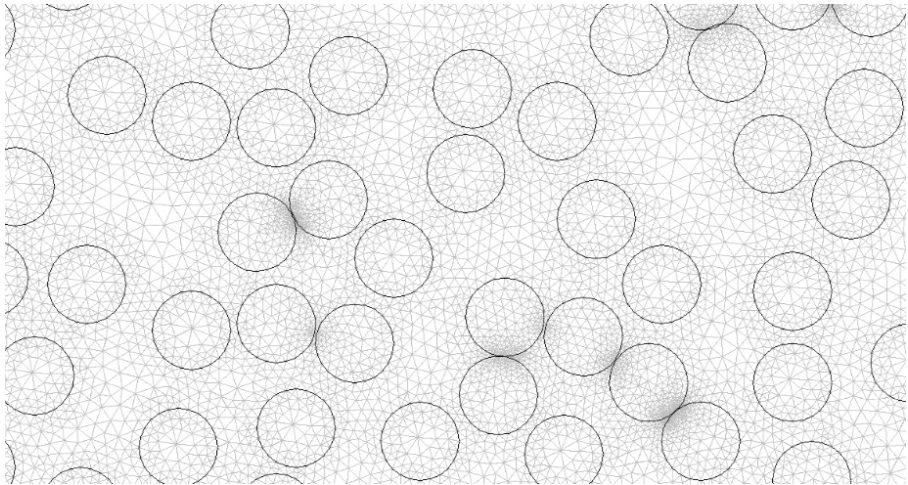}
\caption{Close up on a typical mesh created by Comsol\texttrademark to describe the 2D random system of Fig.~\ref{Fig1_1}.}
\label{FigMesh}
\end{figure}
One of the most important step of the Finite Element Method is the creation of the mesh which describes the system. Figure \ref{FigMesh} shows a close up on a typical mesh calculated for the 2D random system of Fig.~\ref{Fig1_1}. The maximum size of elements must be smaller than 7 times the wavelength \cite{MaxSizeFEM}.

\end{document}